\documentclass[twocolumn,usenames,dvipsnames]{aastex701}

\usepackage{amsmath}
\usepackage{cleveref}

\shorttitle{How Do Disk Galaxies Form?}
\shortauthors{Semenov}
\graphicspath{{./}{figures/}}

\usepackage{etoolbox}
\makeatletter
\patchcmd{\NAT@citex}
  {\@citea\NAT@hyper@{\NAT@nmfmt{\NAT@nm}\NAT@date}}
  {\@citea\NAT@nmfmt{\NAT@nm}\NAT@hyper@{\NAT@date}}
  {}
  {}
\patchcmd{\NAT@citex}
  {\@citea\NAT@hyper@{%
     \NAT@nmfmt{\NAT@nm}%
     \hyper@natlinkbreak{\NAT@aysep\NAT@spacechar}{\@citeb\@extra@b@citeb}%
     \NAT@date}}
  {\@citea\NAT@nmfmt{\NAT@nm}%
   \NAT@aysep\NAT@spacechar%
   \NAT@hyper@{\NAT@date}}
  {}
  {}
\patchcmd{\NAT@citex}
  {\@citea\NAT@hyper@{%
     \NAT@nmfmt{\NAT@nm}%
     \hyper@natlinkbreak{\NAT@spacechar\NAT@@open\if*#1*\else#1\NAT@spacechar\fi}%
       {\@citeb\@extra@b@citeb}%
     \NAT@date}}
  {\@citea\NAT@nmfmt{\NAT@nm}%
   \NAT@spacechar\NAT@@open\if*#1*\else#1\NAT@spacechar\fi%
   \NAT@hyper@{\NAT@date}}
  {}
  {}
\makeatother

\def\Mstar{M_{\rm \star}}
\def\Mb{M_{\rm b}}
\def\Ob{\Omega_{\rm b}}
\def\Om{\Omega_{\rm m}}

\def\SFR{\dot{M}_{\rm \star}}

\def\Mvir{M_{\rm vir}}
\def\rvir{r_{\rm vir}}
\def\vvir{v_{\rm vir}}

\def\cnfw{c}

\def\vc{v_{\rm c}}
\def\vrot{v_{\rm rot}}

\def\xdisk{x_{\rm disk}}
\def\xcirc{x_{\rm circ}}
\def\xturb{x_{\rm turb}}
\def\rcirc{r_{\rm circ}}
\def\rturb{r_{\rm turb}}
\def\rb{r_{\rm b}}

\def\xb{x_{\rm b}}
\def\fb{f_{\rm b}}

\def\jgas{j_{\rm gas}}

\def\kms{{\rm \;km\;s^{-1}}}
\def\Msunyr{{\rm \;M_\odot\;yr^{-1}}}
\def\Msun{{\rm \;M_\odot}}

\def\Myr{{\rm \;Myr}}

\def\cc{{\rm \;cm^{-3}}}
\def\K{{\rm \;K}}

\newcommand{\newtext}[1]{#1}

\begin{document}

\title{How Do Disk Galaxies Form?}

\author[0000-0002-6648-7136]{Vadim A. Semenov}
\email{vadim.semenov@cfa.harvard.edu}
\affiliation{Center for Astrophysics $|$ Harvard \& Smithsonian, 60 Garden St., Cambridge, MA 02138, USA}

\begin{abstract}
In both observed and simulated galaxies, disk morphologies become more prevalent at higher masses and lower redshifts. To elucidate the physical origin of this trend, we develop a simple analytical model in which galaxy morphology is governed by the competition between rotational support and turbulence in a gravitational potential of a dark matter halo and the galaxy itself, and a disk forms when the potential steepens due to the accumulation of baryons in the halo center.
The minimum galaxy mass required for this transition decreases with an increasing dark matter contribution within the galaxy, making more concentrated halos more prone to forming disks. Our model predicts that galaxy sizes behave qualitatively differently before and after disk formation: after disks form, sizes are governed by the halo spin, in agreement with classical models, whereas before disk formation, sizes are larger and set by the scale on which turbulent motions, which dominate over rotation, can be contained.
We validate our model against the results of the TNG50 cosmological simulation and, despite the simplicity of the model, find remarkable agreement. In particular, our model explains the increase with redshift in the critical halo mass for disk formation, reported in both simulations and observations, as a consequence of the evolution of the halo mass--concentration and baryonic mass--halo mass relations. This redshift trend therefore supports the recent proposal that it is the steepening of the gravitational potential that \newtext{primarily drives} disk formation, while other effects discussed in the literature, such as potential deepening and hot gaseous halo formation, can still play important roles in the transition from early turbulent to dynamically cold disks.
\end{abstract}


\section{Introduction}

Disk galaxies account for most of the star formation in the present-day Universe, yet the physical mechanism of their formation remains unclear. Observations and simulations alike show that at early times and low masses, galaxies are typically irregular and dispersion-dominated, while rotationally supported disks become increasingly common at lower redshift and higher masses (see the references below). Understanding when and why galaxies transition from irregular morphologies to ordered disks remains a central problem in galaxy formation theory.
\newtext{Note that the definition of a ``disk'' is ambiguous and depends on the adopted tracers and metrics (see Section~\ref{sec:model}); throughout, we focus on rotation-dominated systems with $\vrot/\sigma \gtrsim 1$, which need not be thin or dynamically cold.}

In the observed population of nearby galaxies \newtext{($z \approx 0$)}, well-developed disks start to dominate over irregular morphologies at stellar masses above $\Mstar \gtrsim 10^9 \Msun$ \citep[e.g.,][]{conselice06,kelvin14,moffett16}. \newtext{This threshold is redshift-dependent and increases toward higher $z$}. The estimated total mass in the pre-disk low-metallicity stellar population of the Milky Way (the so-called ``Aurora'' or ``proto-Galaxy'') of $\sim 5 \times 10^8 \Msun$ in \citet{bk23} and \citet{kurbatov24} suggests that the Milky Way's disk formed at a similar stellar mass threshold, although \citet{conroy22} reports a somewhat lower value, $\sim 10^8 \Msun$. Observations of ionized gas kinematics out to $z \sim 3$ suggest an increasing prominence of dispersion dominated systems due to the increase of turbulent velocities in the interstellar medium (ISM), shifting this mass threshold to higher masses \citep[e.g.,][see \citealt{conselice14} and \citealt{forsterschreiber-wuyts20} for reviews]{kassin12,wisnioski15,wisnioski19,huertas-company16,simons17,turner17}. 

Based on these results, it was long believed that disk galaxies mainly assemble relatively late in the history of the Universe, starting at cosmic noon ($z \sim 2$) and lower redshifts. However, recent observations have revealed a significant population of disk galaxies at much earlier times, $z \sim 4\text{--}8$, which was enabled by the sensitivity of JWST to the rest-frame optical emission from older stellar populations and rest-frame UV at high $z$ \citep[e.g.,][]{ferreira22a,ferreira22b,degraaff24,robertson23,danhaive25} and the sensitivity of ALMA to cold ISM via major cooling lines \citep[such as \ion{C}{2} and \ion{O}{3}; e.g.,][]{smit18,neeleman20,rizzo20,rizzo21,parlanti23,rowland24}. Similarly, the chemo-kinematic data of nearby stars also suggest that the Milky Way's disk formed early, within the first 2 Gyr \citep{bk22,conroy22,rix22,xiang-rix22}. These discoveries imply that disk galaxies were also ubiquitous in the very early Universe. 

The existence of a galaxy mass threshold for disk formation has also been reported in simulations, both large-volume \citep{feng15,pillepich19,trayford19,dubois21,semenov23b,benavides25,celiz25} and high-resolution cosmological runs targeting individual galaxies \citep[e.g.,][]{el-badry18,dekel20,semenov24a}, with disk morphologies dominating in systems with stellar mass $\gtrsim 10^9 \Msun$ or host dark matter halo mass $\gtrsim 10^{11} \Msun$.  This agreement across simulations with very different numerical treatments of gravity, gas cooling, heating, star formation, and feedback indicates that disk formation can be attributed to more general processes that are reasonably well captured in simulations, independent of these details. The physical origin of this transition and its evolution with redshift is the focus of this paper.

A long-standing theoretical understanding is that galactic disks form as a result of the dissipative collapse of gas that retains its angular momentum until it settles into rotational support \citep[e.g.,][]{fall-efstathiou80,kashlinsky82,mo-mao-white98}. In classical models, galaxies share their specific angular momentum with their host dark matter halos, as both are generated by the tidal torques during structure formation \citep[e.g.,][]{doroshkevich70}, leading to disk sizes that scale with the characteristic \newtext{size and} angular momentum of their halos. Such correspondence \newtext{between galaxy and halo sizes} is indeed \newtext{suggested by the observations of} both nearby and high-redshift galaxies \citep[e.g.,][]{kravtsov13,shibuya15,huang17,somerville18}\newtext{, which constrain sizes rather than angular momenta, since the halo spin is not directly observable}, indicating that the classical picture applies, at least on average. 

However, both observations and cosmological simulations suggest that angular momentum alone is not sufficient to predict disk formation. Indeed, galaxies in low-mass halos often fail to form disks even when their specific angular momentum is comparable to that of more massive, disky systems \citep[e.g.,][]{vandenbosch01,sanchez-janssen10,wheeler17}, while simulations do not show a clear correlation between halo spin and galaxy diskiness, size, or spin at fixed mass, as would be expected in such models \citep[e.g.,][]{sales12,desmond17,jiang19,yang23}. 

\newtext{Simulations add a further complication.} Early galaxy simulations that followed structure formation, gas cooling, and dynamics were unable to produce realistic galaxies due to catastrophic losses of angular momentum \citep[the ``angular momentum catastrophe''; e.g.,][]{navarro-white94,navarro95,navarro-steinmetz00}. Although these losses were shown to be in part numerical \citep[e.g.,][]{governato04,governato07,torrey12,vogelsberger12}, the inclusion of efficient stellar feedback appeared to be crucially important for the formation and survival of disks, as feedback can remove low angular momentum gas from galaxies, bring-in high angular momentum from the halo, and keep the disk mass relatively low, ensuring its stability \citep[e.g,][]{okamoto05,scannapieco08,zavala08,brook12,ubler14,agertz15,agertz16,genel15,christensen16,defelippis17}.

Thus, even though the angular momentum of gas is a critically important factor, the physical picture of disk formation is more nuanced than the classical models suggest. 
Over the past decade, several theoretical models have been proposed to explain the transition from dispersion-dominated systems to galactic disks. Broadly, these models attribute disk formation to qualitative changes in either the accretion flow feeding the galaxy or the gravitational potential in which the galaxy evolves.

For example, \citet{dekel20} attributed the mass threshold for disk formation to the halo mass scale at which mergers with misaligned angular momentum, causing spin flips of the disk, become less frequent. In this model, the potential also plays a role: it must be deep enough to prevent stellar feedback from significantly perturbing the ISM, thereby enabling the galaxy to settle into a disk. Apart from destructive mergers, the formation of a disk can depend strongly on the configuration of the accretion flow, such that a substantial inflow with misaligned angular momentum can drive significant velocity dispersions and lead to variations in the galaxy spin axis direction, preventing a disk from settling \citep[e.g.,][]{sales12,meng-gnedin21,semenov23b}. While mergers and the geometry of accretion flows are undoubtedly important for the formation and survival of individual disks, in this paper, we focus on the average behavior and seek the conditions for dark matter halos and galaxies themselves required to sustain such disks.

Another change in the accretion flow relevant for galaxy morphology is the formation of a hot gaseous halo and the associated transition from cold to hot mode accretion \citep[e.g.,][]{dekel-birnboim06,dekel09}. At halo masses lower than $\Mvir < 10^{12} \Msun$, the infalling gas can cool efficiently and form cold, supersonic accretion streams that connect the galaxy to the surrounding cosmic web and, depending on its geometry, can feed the galaxy by misaligned angular momentum and drive large velocity dispersions in the ISM due to large infall velocities of the order of $\vvir$ (the cold-mode accretion). At larger masses, when the virial temperature becomes $T_{\rm vir} \gtrsim 10^6\K$ and the gas is heated to such temperatures through accretion shocks or feedback, its cooling efficiency reduces substantially. \citet{stern19} proposed that this transition can explain the settling of galaxies from highly bursty early evolution to dynamically cold disks. Specifically, in such a hot mode, the accretion flow becomes subsonic and joins the ISM more gradually, driving lower velocity dispersions and coherently delivering angular momentum, as misaligned contributions can be canceled out during the inflow \citep[see also][]{stern20,stern21,stern23,gurvich22,hafen22}.
The cold-to-hot mode transition, however, occurs at significantly higher masses than those at which galactic disks are found in both observations and simulations ($\Mvir \sim 10^{11}\Msun$ \newtext{at low redshift}; see the references above), implying that this transition cannot fully explain formation of disks but it can be important in their subsequent settling from highly turbulent, thick to dynamically cold, thin disks (see Section~\ref{sec:discussion:timeline}).

In addition to changes in the accretion flow, changes in the gravitational potential can also play a critical role in disk formation. For example, as a prerequisite for disk formation, the potential well needs to be deep enough to contain the gas motions imparted by feedback \citep[][]{dekel86,maclow99-blowout,ferrara00,read06}. Further deepening the potential reduces fluctuations induced by individual feedback events, enabling disks to settle \citep[e.g.,][]{el-badry18,dekel20,hopkins23disk}. Understanding disk formation in such a scenario then requires understanding the coupling between stellar feedback and ISM turbulence, which is highly uncertain, especially at low masses and extreme conditions in the early Universe, when disks are not yet established. Additionally, the relative contributions of stellar feedback and gravity in driving turbulence remain debated \citep[e.g.,][]{krumholz16}.

Recently, \citet{hopkins23disk} proposed that disk formation is triggered not by the deepening of the potential but rather by developing a central mass concentration that steepens the potential. The authors conducted a series of galaxy formation simulations with varying parameters, including the depth and shape of the gravitational potential and the efficiencies of gas cooling and feedback. They concluded that, while it is possible to produce non-disk morphologies even in deep potentials, the presence of the disk is strongly correlated with the steepness of the potential. The authors attributed this connection to a steep potential that provides a well-defined geometric center for disk formation and efficiently damps radial motions of the gas (breathing modes). In the context of cosmological galaxy evolution, however, the causal connection remains debated, as the formation of a disk can itself cause steepening \citep[][]{dillamore24,semenov23b}. In addition, the existence of bulgeless thin disk galaxies \citep[][]{kormendy10,kormendy12,fisher11,bizyaev21,haslbauer22} challenge such a requirement of a central mass concentration for disk formation \citep[although see the discussion in][]{hopkins23disk}. It is also interesting that this model was supported by the results from FIRE simulations, which were also argued to support the hot halo formation scenario outlined above \citep[][]{stern21,gurvich22,hafen22}, indicating that establishing causal connections can be challenging even in numerical simulations. 

In this work, we use analytical arguments and cosmological simulations to elucidate the origin of the mass threshold for disk formation and its redshift dependence. 
To this end, in Section~\ref{sec:model}, we develop a simple analytical framework to describe the above models and, based on it, a criterion for disk formation. In Section~\ref{sec:results}, we test our model against the large-volume cosmological simulation TNG50 and show that, despite its simplicity, the model describes disk formation in this simulation remarkably well. To avoid the causal uncertainty related to possible effects of disk formation on their dark matter halos, we specifically use the halo properties from the dark-matter-only version of TNG50 as an input to our model and predict disk formation in the fiducial run. 
We discuss our results in Section~\ref{sec:discussion} and summarize our conclusions in Section~\ref{sec:summary}.

\section{Analytic model for disk formation}
\label{sec:model}

Whether a galaxy settles into a disk depends on the relative importance of two types of motion competing against gravity: coherent rotation and turbulence. 
For this reason, the prominence of a disk is often quantified via the ratio of rotational velocity to velocity dispersion, $\vrot/\sigma$. This ratio and the existence of a disk itself can be defined in different components, e.g., gas or stars, with a further subdivision to different ISM phases (e.g., neutral \ion{H}{1} and cold molecular disks or disks probed by the ionized gas) and stellar populations (e.g., young or old stellar disks). Given that the existence of a long-lived gaseous disk is a prerequisite for the build-up of a stellar disk, we will focus on the formation of gaseous disks. 

Thus, the goal of this work is to understand when and why, in the evolution of a galaxy, the transition from the early, dispersion-dominated ISM states with $\vrot/\sigma \lesssim 1$ at low masses to dynamically cold gaseous disks with $\vrot/\sigma \gg 1$ at high masses occurs. 
\newtext{In this section}, we develop a criterion linking galaxy morphology to the basic properties of its dark matter halo, which can be predicted by large-scale structure formation models, such as halo mass, concentration, and spin parameter. \newtext{To help interpret this model, we place it in the context of other proposed disk-formation scenarios in Section~\ref{sec:discussion:models} and summarize the resulting timeline of disk assembly in Figure~\ref{fig:disk-timeline}.}

\subsection{Connection between Gravitational Potential and Disk Formation}
\label{sec:model:idea}

\begin{figure}
\centering
\includegraphics[width=\columnwidth]{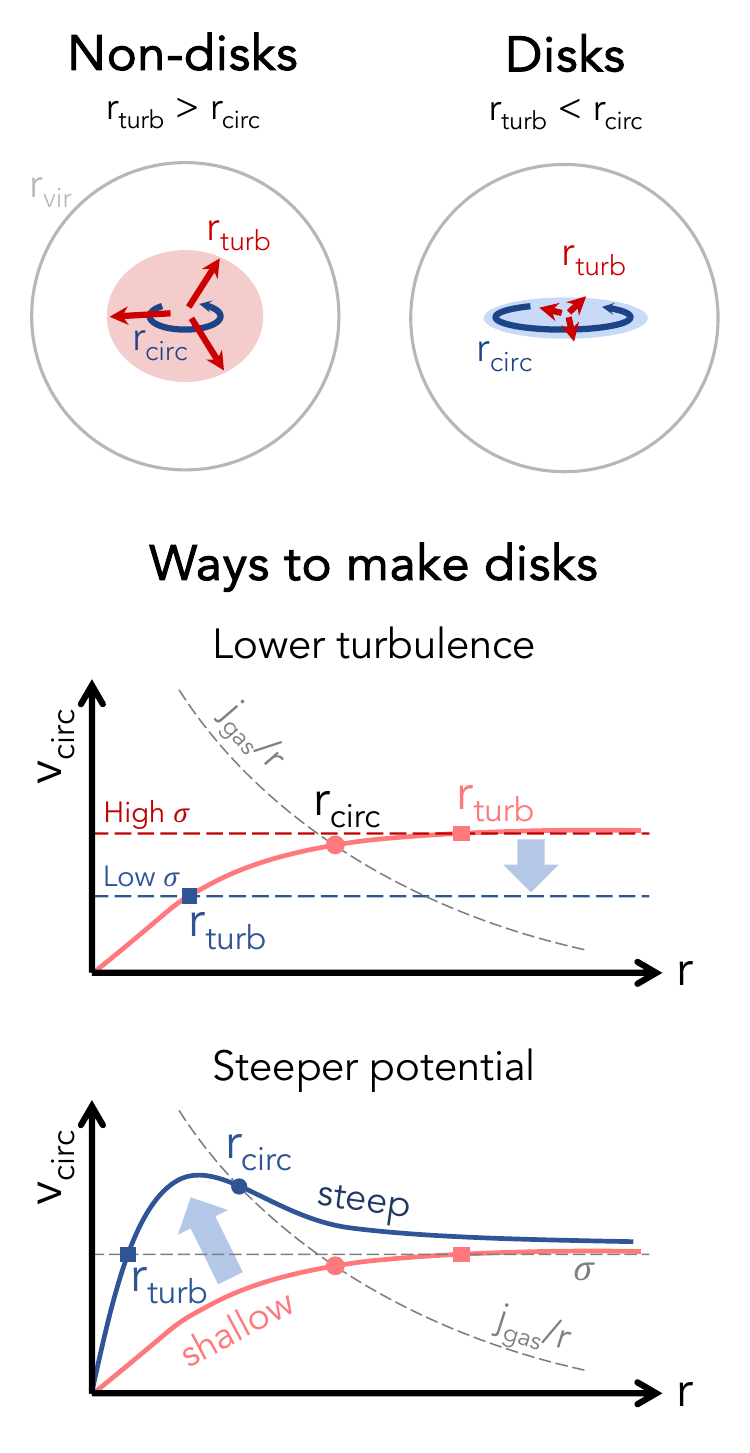}
\caption{\label{fig:schematics} Schematic illustration of the disk formation scenarios and the role played by the gravitational potential (or the shape of the rotation curve). {\bf Top:} a disk forms when the scale on which baryons settle into rotational support, $\rcirc$, is larger than the scale on which the turbulent motions are contained by the gravitational potential, $\rturb$. With the definitions described in the text, this criterion, $\rcirc/\rturb > 1$, is equivalent to $\vrot/\sigma > 1$. {\bf Bottom:} the transition from non-disks ($\rcirc/\rturb < 1$) to disks ($\rcirc/\rturb > 1$) can be achieved either by decreasing $\sigma$ relative to $\vvir$, or by making the potential steep, resulting in a peaked rotation curve.}
\end{figure}

\subsubsection{Why Does the Shape of the Potential Matter?}
\label{sec:model:idea:general}

The three key factors determining whether a given galaxy would form a rotationally supported gaseous disk with high $\vrot/\sigma$ are the specific angular momentum carried by the gas, $\jgas$, the driver of turbulence setting $\sigma$, and the shape of the gravitational potential, $\phi(r)$. The latter sets both the shape of the rotation curve, $\vc^2(r) = r d\phi/dr \approx GM(<r)/r$ (assuming spherical symmetry or in the disk plane), and the escape velocity, $v^2_{\rm esc} (r) = 2 |\phi| \approx 2 GM(<r)/r \approx 2\vc^2(r)$, thereby determining whether rotation or turbulence dominates the support and setting the characteristic size of the resulting system.

This connection is schematically illustrated in Figure~\ref{fig:schematics}. For given $\jgas$ and $\sigma$, the shape of the potential determines two spatial scales on which this angular momentum and turbulence can be contained, $\rcirc$ and $\rturb$, respectively.
If $\jgas$ is sufficiently high, the gas settles into a rotationally supported disk with the size, $\rcirc$, and rotational velocity, $\vrot$, connected by $\vc(\rcirc) = \jgas / \rcirc \equiv \vrot$. Alternatively, if $\sigma$ is large, turbulence disrupts the gas before rotational support is established, and the gas joins the dispersion-dominated motions on the scale of $\rturb > \rcirc$. The scale $\rturb$ is set by the depth of the potential well (via $v_{\rm esc}$) and, up to order-unity factors, can be estimated as the radius at which $\vc(\rturb) \sim \sigma$. A more accurate estimate depends on the details of mass distribution, the exact definition of the size of the system, and the interplay between turbulent support and gravity (see also Section~\ref{sec:results:mw-analogs}), but as our results below show, this simple definition of $\rturb$ is sufficient for our purpose of formulating a disk formation criterion. This is because such a definition captures the rapid decrease of $\rturb$ in steep potentials.

It is worth noting that both $\jgas$ and $\sigma$ can vary stochastically, especially during the early stages of disk formation, when active accretion and frequent mergers dominate, leading to the transient nature of such disks at low galaxy masses and early times. Indeed, the direction of accreted angular momentum can change drastically, leading to frequent disk reorientation and episodes of its formation and destruction \citep[see also][]{sales12,dekel20,kretschmer22,meng-gnedin21}. Similarly, even if a disk can form, mass blowouts caused by stellar feedback can drive violent fluctuations of $\sigma$, preventing disk settling and causing the above conditions for disk formation to appear and disappear repeatedly \citep[e.g.,][]{agertz16,yu22,gurvich22}. In this paper, we seek to derive the average conditions required for disk formation, and therefore assume that $\jgas$ and $\sigma$ change smoothly across this transition. Accretion of misaligned material, mergers, and feedback-induced fluctuations of $\sigma$ will produce galaxy-to-galaxy variations around these conditions (see Section~\ref{sec:discussion:disks}).

It is also important to note that to survive after formation, the disk must remain sufficiently low mass to avoid violent instabilities that can rapidly turn it into a spheroid \citep[e.g.,][]{fall-efstathiou80,agertz16}. Stellar feedback, therefore, plays another critical role in disk formation and survival by keeping its mass fraction low. In this paper, we only derive the conditions for disk formation and do not explicitly consider the requirements for its stability, assuming that stellar feedback is sufficiently efficient to ensure disk survival (see also Section~\ref{sec:model:idea:assumptions}). We further discuss the multifaceted roles of feedback in disk formation and how they are incorporated into our model in Section~\ref{sec:discussion:feedback}. 

The above definitions of the scales of rotational support, $\rcirc$, and turbulence confinement, $\rturb$, can be used to both introduce a formal criterion for disk formation and determine the average galaxy size, as the latter is set by the largest of the two scales. Indeed, if $\rcirc < \rturb$, a disk with the size $\rcirc$ cannot exist in equilibrium as its high velocity dispersion $\sigma$ would lead to an expansion until the characteristic size becomes comparable to the scale on which this $\sigma$ can be contained, $\rturb$. From the momentum conservation, the rotational velocity of the galaxy reaches $\vrot \sim \jgas/\rturb < \sigma$ (see Figure~\ref{fig:schematics}). Similarly, if $\rturb < \rcirc$, a galaxy will settle into a rotationally supported disk with the size $\rcirc$ and $\vrot = \jgas / \rcirc = \vc(\rcirc) > \sigma$. In other words, with these definitions, the condition $\rcirc / \rturb > 1$ is equivalent to $v_{\rm rot}/\sigma > 1$.

In general, there are two qualitatively different ways in which the transition from non-disks ($\rturb > \rcirc$) to disks ($\rturb < \rcirc$) can occur, as also shown in Figure~\ref{fig:schematics}. First, $\sigma$ can decrease (or, equivalently, $\jgas$ can increase) until $\rturb$ becomes smaller than $\rcirc$, reflecting the evolution of the turbulent state of the galaxy and/or the characteristics of the accretion flow (top panel in the lower set). Second, the shape of the potential can change, so that for fixed $\sigma$ and $\jgas$, a disk can develop in a sufficiently steep potential with a peaked $\vc(r)$ even if the velocity dispersion is too high for such a disk to form in a shallow potential (bottom panel).

This picture helps to understand the qualitative differences between the scenarios for disk formation outlined in the Introduction. To this end, it is first convenient to normalize $\vc$ and $r$ by $\vvir$ and $\rvir$, respectively, and assume that the specific angular momentum of baryons follows that of the dark matter halo, $\jgas \sim \lambda \rvir \vvir$, with the halo spin parameter $\lambda$ independent of the halo mass (see Section~\ref{sec:model:idea:assumptions}). With these assumptions, the line of constant angular momentum in Figure~\ref{fig:schematics} stays fixed, and all the changes in these scenarios can be thought of as either the changes of $\sigma$ relative to $\vvir$ or the changes in the shape of $\vc(r)$. 

\emph{Containment of feedback-driven motions by the halo potential} \citep{dekel86}: As a prerequisite for disk formation, the motions imparted into the gas by feedback must remain bound in the potential provided by the dark matter halo and baryons \citep[see also][]{maclow99-blowout,ferrara00,read06}. Based on rough arguments, \citet{dekel86} estimated $\sigma \sim \sigma_{\rm fb} \sim 100\kms$, but the exact value and its possible dependence on halo mass and redshift depends on the uncertain details of the feedback energy coupling to the ISM gas. A gas disk can settle only when the halo is massive enough to contain these motions, requiring that $\vvir \equiv \sqrt{G\Mvir/\rvir} > \sigma_{\rm fb}$. Note, however, that this is a necessary condition; the sufficient condition further requires that $\sigma_{\rm fb} < \vrot$, which depends on $\sigma_{\rm fb}$, $\jgas$, and the shape of $\vc(r)$. In the simple schematics of Figure~\ref{fig:schematics}, this scenario is equivalent to sliding the horizontal line of $\vc/\vvir = \sigma/\vvir$ down due to the increasing $\vvir$, analogous to the first scenario outlined above. 

\emph{Development of the inner hot halo} \citep{stern23}: in this scenario, velocity dispersion is set by gas accretion via the cooling flow, $\sigma \sim \sigma_{\rm acc}$. In the cold-mode accretion at low masses, $\Mvir < 10^{12} \Msun$, gas cools efficiently out to large radii $\sim \rvir$ and then falls supersonically on the galaxy. As a result, gas is delivered to the galaxy in cold streams that are accelerated by the gravitational potential and can entrain a significant mass from the surrounding halo gas \citep[e.g.,][]{aung24}, thereby imparting large momentum and driving high velocity dispersions of the order $\sigma_{\rm acc} \sim \vvir$ especially in low-mass systems. At around $\Mvir \sim 10^{12} \Msun$, a hot gaseous halo develops as a result of accretion shock and feedback heating to halo virial temperature \citep[][]{birnboim-dekel03,dekel-birnboim06,dekel09}. This slows down the inflow to subsonic velocities, enabling the gas to join the galaxy smoothly with $\sigma_{\rm acc} < \vvir$. Similar to the above, this scenario is equivalent to sliding the $\vc = \sigma_{\rm acc}$ line down to the point when $\sigma_{\rm acc}$ becomes $ < \vrot$, enabling disk formation.

\emph{Steepening of the inner potential} \citep{hopkins23disk}: The above two scenarios imply that disk settling, i.e., $\rturb < \rcirc$, results from the relative change of $\sigma/\vvir$ at large $\Mvir$ without changing the shape of the potential. A qualitatively different way to satisfy $\rturb < \rcirc$ is by developing a sufficiently steep potential with a peaked rotation curve as illustrated in the bottom panel of Figure~\ref{fig:schematics}. This mechanism can describe disk formation even in the low-mass regime with large $\sigma \sim \vvir$, corresponding to the formation of thick, highly-turbulent disks with $\vrot/\sigma \gtrsim 1$. This is one of the motivations for considering this scenario in our disk-formation criterion below.

\subsubsection{Our Model Assumptions}
\label{sec:model:idea:assumptions}

In our model, we consider the evolution of a galaxy from a very early stage without a gaseous disk, $\vrot/\sigma < 1$, and seek to identify the conditions when a disk with $\vrot/\sigma > 1$ forms. As described above, it is convenient to factor out the halo mass dependence by normalizing all velocities and spatial scales by $\vvir$ and $\rvir$, respectively. In this case, the above scenarios of disk formation require us to specify three dimensionless parameters describing turbulence, $\sigma/\vvir$, specific angular momentum $\jgas/\rvir\vvir$, and the shape of the rotation curve, $\vc(x)/\vvir$, where $x \equiv r/\rvir$.
Our strategy is to derive a disk formation criterion within the space of these dimensionless parameters (Sections~\ref{sec:model:profile}--\ref{sec:model:criterion}) and then investigate the dependence of these parameters on halo mass and redshift to elucidate disk formation in cosmological simulations (Section~\ref{sec:results}).

To quantify rotational support, we assume that gas inherits its angular momentum from the large-scale structure formation and retains it on average \citep{fall-efstathiou80,mo-mao-white98}. In this case, angular momentum conservation implies that the final disk size is set by the size and the spin parameter of the dark matter halo: $\vrot\,\rcirc \sim \jgas \sim j_{\rm dm} \sim \lambda \rvir \vvir$, so that $\rcirc \sim \lambda \rvir (\vrot / \vvir)^{-1}$ with $\vrot = \vc(\rcirc)$. 
For realistic matter distributions, $\vrot/\vvir$ stays within a factor of $\sim 2$, leading to the classical result $\rcirc \sim \lambda \rvir$, with correction factors depending on the details of the matter distribution and the degree of angular momentum retention \citep[][]{mo-mao-white98}. 
As described in the Introduction, this relation holds on average for nearby and high-redshift galaxies, making the assumption of angular momentum retention a natural approximation for our purposes. As also noted above, we further assume that the angular momentum is, on average, accreted smoothly and coherently; deviations from this assumption due to mergers and misaligned accretion result in a galaxy-to-galaxy variation of the disk formation threshold described here (see also Section~\ref{sec:discussion:disks}).  

In terms of the schematics shown in Figure~\ref{fig:schematics}, these assumptions effectively fix the location of the constant angular momentum line with the normalization set by $\lambda$, and the smallest irreducible galaxy size set by the full rotational support, $\rcirc/\rvir \sim \lambda$. The distribution of dark matter spin parameter, $\lambda$, is one of the robust predictions of cosmological structure formation simulations, which produce a log-normal distribution with the mean value of $\lambda \approx 0.035$ and $\sigma_{\log \lambda} \approx 0.22$ dex scatter \citep[e.g.,][see \citealt{mo-vandenbosch-white} for a review]{bett07,maccio07,yang23}.

Next, as we seek the minimal halo (or galaxy) mass threshold at which condition $\vrot/\sigma > 1$ can be satisfied, we assume the maximal possible value of the velocity dispersion that a given halo can contain: $\sigma \sim \vvir$. \newtext{This choice is a simplification to which our disk formation criterion is only weakly sensitive, as we discuss below in this subsection.} In halos with $\vvir < \sigma$, turbulent motions would be unbound, and the galaxy would not form. In general, the condition $\sigma \leq \vvir$ should be required in addition to the criterion that we are deriving here, however, as we show below, the evolution of the disk formation threshold with redshift suggests that disk formation is triggered by the change of the potential rather than the evolution of $\sigma/\vvir$ (see Sections~\ref{sec:results:disk-population} and \ref{sec:discussion:models}). The latter can be important at very low $\vvir$ and in the presence of strong feedback, which can drive strong turbulence in early disks, making it challenging to distinguish them from non-disky systems (see Sections~\ref{sec:discussion:timeline} and \ref{sec:discussion:feedback}). Our criterion, therefore, should be thought of as the minimal halo mass that can sustain galactic disks in the limit of high $\sigma \sim \vvir$.
In Figure~\ref{fig:schematics}, this assumption effectively fixes the location of the horizontal dashed line at $\sigma/\vvir = 1$. We further assume that the galaxy size in this regime is approximately equal to $\rturb$, at which $\vc(\rturb) = \sigma = \vvir$.

With these assumptions, the locations of constant $\sigma$ and constant angular momentum lines in Figure~\ref{fig:schematics} (dashed lines) are both fixed, and, as $\vc(r)/\vvir \to 1$ at $r/\rvir \to 1$, the \newtext{primary} way to trigger disk formation is by developing a sufficiently steep potential in the center \citep[the bottom panel in the figure;][]{hopkins23disk}. We therefore consider such steepening as the main driver of disk formation at the threshold when the condition $\vrot/\sigma \sim 1$ is met in the evolution of a galaxy for the first time.
As the formal criterion of when the potential is sufficiently steep, we require that $\vc$ at the scale of the galaxy size equals $\vvir$, implying that $\vc(r)$, $\sigma = \vvir$, and $\jgas = {\rm const}$ all intersect in the same point in Figure~\ref{fig:schematics}. In what follows, we define the galaxy size as the radius that encloses half of its baryon mass (gas and stars).

The assumption that $\sigma \sim \vvir$ at the disk formation stage is admittedly a simplification. It is motivated by the fact that at early times, the infalling gas can cool efficiently out to the virial radius, bringing in significant mass and momentum \citep[e.g.,][]{aung24} and thereby imparting velocity dispersions of the order of the characteristic virial velocity in the halo center. A more accurate model for $\sigma$ can be derived by considering the energetics of gas accretion and turbulence dissipation \citep[e.g.,][]{elmegreen-burkert10}. Note, however, that the disk formation criterion is only weakly sensitive to the exact value of $\sigma$ because, as the rotation curve develops a peak, the scale on which turbulent velocities are confined, $\rturb$, decreases rapidly for any value of $\sigma$ (see the bottom panel of Figure~\ref{fig:schematics}). However, a more accurate model for $\sigma$ and $\rturb$ is important for predicting the sizes of early galaxies before they develop a disk and properties of early turbulent disks with $\vrot/\sigma \gtrsim 1$ (see also Section~\ref{sec:discussion:sizes}). We leave this to a follow-up study.

Finally, as noted above, stellar feedback plays a central role in disk formation. First, feedback regulates the growth of the central baryonic concentration, which causes the steepening of the potential. Second, after disk formation, feedback ensures that its mass stays sufficiently low to avoid violent instabilities and enable disk survival. Third, it regulates the distribution of angular momentum in the galaxy and halo, preventing the accumulation of low-angular momentum gas in the center. Finally, feedback contributes to stochastic changes in turbulent velocities, introducing variations in the timing of disk formation and its subsequent settling (see Section~\ref{sec:discussion:timeline}). We further discuss how these different roles of feedback enter our model in Section~\ref{sec:discussion:feedback}.

Our main assumptions can be summarized as follows:

\begin{itemize}
    \item In the lowest mass halos that can sustain disks, ISM turbulence is primarily driven by gas accretion in cold mode, leading to $\sigma \sim \vvir$.

    \item Disk forms when the potential becomes steep enough, so that $\vc$ at the galaxy half-baryon-mass radius, $\rb$, becomes $\vc(\rb) > \sigma \sim \vvir$, which is equivalent to $\rcirc/\rturb > 1$ in Figure~\ref{fig:schematics}.

    \item For disks, the galaxy size is set by the rotational support radius, determined by the halo size and spin, assuming angular momentum retention, $\rcirc \sim \lambda \rvir$.

    \item For non-disks, the galaxy size is $\rturb > \rcirc$, i.e., the scale on which the turbulent motions with $\sigma > \vrot$ are contained.
\end{itemize}
These assumptions enable one to determine whether a galaxy within a given potential (i.e., $\vc(r)$) will have a disk or not. The remaining piece needed to derive a disk formation criterion is the distribution of matter and its dependence on the halo mass, setting $\vc(r)$.

\subsection{The Model for Matter Distribution}
\label{sec:model:profile}

We model the potential by assuming a Navarro--Frenk--White (NFW) dark matter halo \citep{navarro96,navarro97} with concentration $\cnfw$ and a two component baryon contribution: a centrally concentrated part, i.e., the galaxy, that contains $\fb$ fraction of all available baryons in the center, with $\rb$ being its half-mass radius, and the rest $(1-\fb)$ in a diffuse halo which is assumed to follow the same profile as the dark matter halo. 
When the radial coordinate and $\vc$ are normalized by $\rvir$ and $\vvir \equiv \sqrt{G\Mvir/\rvir}$, the rotation curve $\vc(r)$ in our model is fully described by three parameters: $\cnfw$, $\fb$, and $\xb \equiv \rb/\rvir$.
The following subsections provide further details about the contributions of different components to our model $\vc(r)$.

\subsubsection{Dark Matter and Diffuse Gas Halo}

\begin{figure}
\centering
\includegraphics[width=\columnwidth]{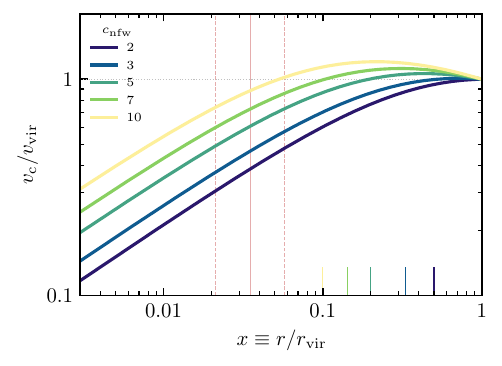}
\caption{\label{fig:vc-nfw} Rotation curves of an NFW dark matter halo with different concentrations $\cnfw$. The corresponding scale-radius, $r_{\rm s}/\rvir = 1/\cnfw$, is shown with the vertical ticks. As a proxy for the galaxy size, the red vertical lines show the range of halo spin parameters, $\lambda$, with the mean value of 0.035 and 0.22 dex of scatter. For a pure dark matter halo, $\vc/\vvir$ evaluated at $x \sim \lambda$ is usually $< 1$, implying shallowly rising potential and therefore a dispersion-dominated system ($\vrot/\sigma < 1$, or $\rcirc/\rturb < 1$).}
\end{figure}

For the dark matter and diffuse baryon components, we adopt an NFW density profile:

\begin{equation}
    \rho_{\rm nfw} = \frac{\rho_0}{(r/r_{\rm s}) (1+r/r_{\rm s})^2},
\end{equation}
where the scale radius, $r_{\rm s}$, is related to the virial radius via the concentration parameter $\cnfw \equiv \rvir/r_{\rm s}$, and $\rvir$ is the radius enclosing the overdensity, $\Delta_{\rm vir}(z)$, for which we adopt the \citet{bryan-norman98} fitting formula. In the rest of the section, we normalize the quantities to their values at $\rvir$ and use $x \equiv r/\rvir$ as the radial coordinate, so that the scale radius corresponds to $x = 1/\cnfw$.

The total mass enclosed within a given radius $r = x\,\rvir$ is given by

\begin{equation}
    M(<r) = \Mvir \frac{g(\cnfw\,x)}{g(\cnfw)},
\end{equation}
where $\Mvir \equiv (4 \pi /3)\,\Delta_{\rm vir} \rho_{\rm crit} \rvir^3$ is the virial mass and
\begin{equation}
    g(x) \equiv \ln(1+x) - \frac{x}{1+x}.
\end{equation} 
Accordingly, the rotational velocity normalized by the virial velocity is given by
\begin{equation}
\label{eq:vc-nfw}
    \frac{v_{\rm c,nfw}(r)}{\vvir} = \left[ \frac{g(\cnfw\,x)}{x\,g(\cnfw)} \right]^{1/2}.
\end{equation}

Overall, a dark matter halo in our model is fully characterized by $\Mvir$ (or, equivalently, $\rvir$ or $\vvir$), $\cnfw$, and $\lambda$, and our ultimate goal is to identify a disk formation criterion using these three numbers.
Figure~\ref{fig:vc-nfw} shows $\vc(r)$ for a representative set of concentrations, with the vertical red lines showing the mean value and scatter of the spin parameter $\lambda$, as a proxy for the galaxy size. According to our model, a disk forms when $\vc/\vvir > 1$ at the scale corresponding to the galaxy size (see the second bullet point in Section~\ref{sec:model:idea:assumptions}).
As the figure shows, a pure NFW halo generally provides a gradually rising potential, with $\vc/\vvir < 1$ at $\lambda \rvir$, and it is therefore unable to sustain a galactic disk, unless its concentration is very high (see also Section~\ref{sec:discussion:disks}).
However, a sufficiently steep potential can be produced by the baryons, which, unlike dark matter, can cool efficiently and concentrate in the halo center.

\subsubsection{Contribution of Baryons to the Central Potential}

We model the contribution of baryons by assuming that a fraction $\fb = (\Mb/\Mvir)/(\Ob/\Om)$ of available baryons is concentrated in the center with the normalized half-mass radius $\xb$. Even for $\fb \sim 1$, the total mass of baryons is subdominant (and equals the universal baryon fraction $\Ob/\Om \approx 0.16$ of $\Mvir$, assuming Planck XIII (\citeyear{planck-xiii}) cosmology), but for sufficiently small $\xb$, it can substantially steepen the gravitational potential in the center and produce a peaked rotation curve. 

\begin{figure}
\centering
\includegraphics[width=\columnwidth]{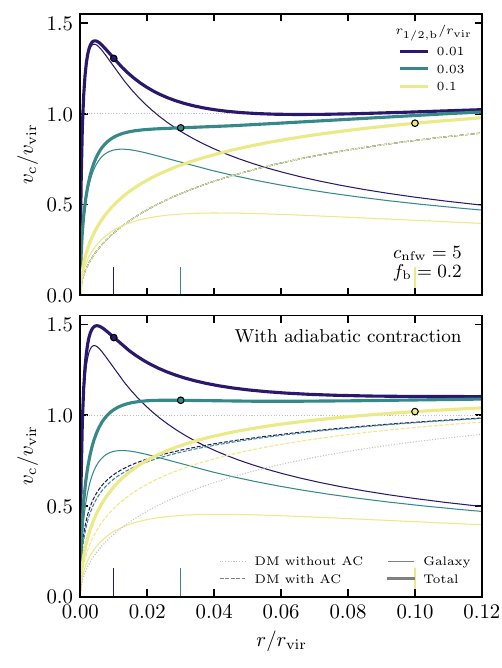}
\caption{\label{fig:vc-bar-ac} The contribution of baryons (including stars and gas) to the gravitation potential in the halo center can produce a sufficiently steep potential to trigger disk formation. Different colors show rotation curves assuming that the $\fb = 0.2$ fraction of available baryons is concentrated at the center, with three different half-mass radii ($\xb$; also shown with the vertical ticks), with the smaller $\xb$ leading to more peaked rotation curves. Different line styles show the total (thick) and individual contributions of baryons (thin) and dark matter (dashed). The top panel shows $\vc(r)$ assuming that the dark halo remains fixed (with the concentration $\cnfw = 5$, which is typical for $\Mvir \sim 10^{11} \Msun$ at $z \sim 2$), while the bottom panel also accounts for the adiabatic contraction of the halo (AC; see Section~\ref{sec:model:profile:ac}). The markers show the value of $\vc/\vvir$ at $\xb$. According to our criterion ($\vc(\xb)/\vvir > 1$), without AC, only the smallest of $\xb$ can produce a steep enough potential to trigger disk formation, while with AC, all three values satisfy this condition. }
\end{figure}

For example, the top panel of Figure~\ref{fig:vc-bar-ac} shows $\vc(r)$ for fixed $\fb = 0.2$ and $\cnfw = 5$ and several values of $\xb$. For $\xb = 0.1$ and 0.03, the value of $\vc/\vvir$ at $\xb$ (indicated with the circle) is still below 1, implying that a disk cannot form according to the above criterion. However, $\xb = 0.01$ is compact enough to ensure $\vc/\vvir > 1$ and therefore trigger disk formation.

In this figure and the rest of the paper, we assume the \citet{hernquist90} profile:
\begin{equation}
    \rho_{\rm b} = \frac{\Mb}{2\pi a^3} \frac{1}{(r/a) (1+r/a)^3},
\end{equation}
It describes stellar bulges and elliptical galaxies and is therefore well suited to our purpose of investigating the onset of disk formation, as such galaxies, by definition, do not have an established disk. Note also, that most of the subsequent discussion is independent of the choice of the profile for the baryonic component, as it is parametrized via the half-mass radius $\xb$ and the value of $\vc/\vvir$ at that radius, which are independent of the choice of the profile (as, by definition, $\xb$ always include exactly a half of the baryon mass, $0.5 \fb\, \Ob/\Om\, \Mvir$). Only the shape of $\vc/\vvir$ around $\xb$ depends on the profile. \newtext{The same holds for the dark matter halo: as the analysis depends only on the resulting $\vc(r)$, adopting a broadly similar profile, such as Einasto instead of NFW, would change $\vc(r)$, and therefore our criterion, only weakly.}

For a Hernquist profile, the total mass enclosed within $r$ is
\begin{equation}
    M(<r) = \Mb \frac{(r/a)^2}{(1+r/a)^2},
\end{equation}
and the half-mass radius is 
\begin{equation}
    \rb = (1+\sqrt{2})\,a.
\end{equation}
Its contribution to the rotation curve is
\begin{equation}
    \frac{v_{\rm c,b}(r)}{\vvir} = \left[ \fb \frac{\Ob}{\Om} \frac{(r/a)^2}{(r/\rvir)(1+r/a)^2} \right]^{1/2},
\end{equation}
which is shown in Figure~\ref{fig:vc-bar-ac} and added to that of the NFW profile (Equation~(\ref{eq:vc-nfw})) in quadrature. For simplicity, we assume that the remaining baryon mass follows the dark matter distribution, and therefore, the contribution of the NFW profile is adjusted by $1-\fb\, \Ob/\Om$ to ensure mass conservation.

\subsubsection{Response of the Dark Matter Halo to Baryons}
\label{sec:model:profile:ac}

Apart from the direct contribution of baryons to the potential, the deeper potential at the center also leads to the accumulation of dark matter, further deepening the potential---the effect known as ``adiabatic contraction'' of the dark matter halo.
We incorporate this effect in our model following \citet{gnedin04}.

The bottom panel of Figure~\ref{fig:vc-bar-ac}, shows the effect of adiabatic contraction on the rotation curves. The additional steepening of the potential is sufficient to ensure $\vc/\vvir > 1$ for all three values of $\xb$, formally enabling disk formation in all three cases, while without this effect, only $\xb = 0.01$ ensured a sufficiently steep potential (see the top panel).

\newtext{Note, however, that the strength of this contraction is uncertain, and some studies suggest that it can be significantly weaker \citep[e.g.,][]{dutton07,abadi10}. In addition,} efficient and sufficiently variable feedback can produce the effect opposite to adiabatic contraction via heating up of the inner dark matter density concentration induced by the potential fluctuations due to repeated mass blowouts \citep[e.g.,][]{governato10,pontzen14,dutton19}. These effects can be especially important for the formation of high-redshift disks due to the increased variability of star formation and feedback at high redshifts suggested by the JWST observations of the early abundances of bright galaxies \citep[e.g.,][]{mason23,sun23-fire,kravtsov-belokurov24}. These effects warrant a further investigation in a separate study (see also Section~\ref{sec:discussion:feedback}).

\subsection{Transition from Turbulence-dominated to Disk Galaxies}
\label{sec:model:transition}

\begin{figure}
\centering
\includegraphics[width=\columnwidth]{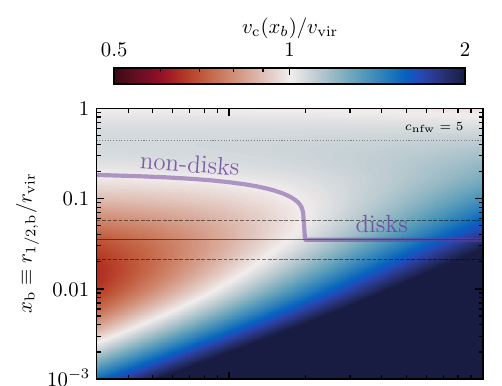}\\
\includegraphics[width=\columnwidth]{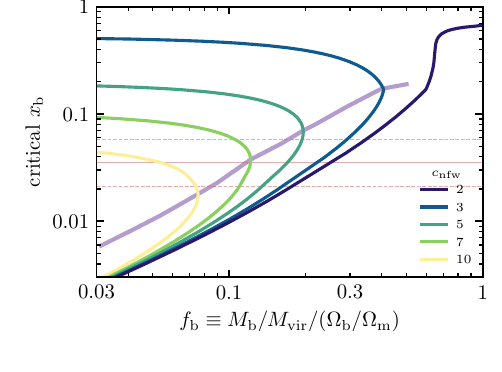}
\caption{\label{fig:fbxb-model} Distinction between disks and non-disks in the plane of the galaxy baryon-half-mass size, $\xb$, and mass fraction, $\fb$. {\bf Top:} The potential steepness parameter ($\vc(\xb)/\vvir$) as a function of $\xb$ and $\fb$, for an NFW halo with concentration $\cnfw = 5$. For reference, the dotted line shows the half-mass radius of the NFW profile, while solid and dashed black horizontal lines show the typical values of the halo spin parameter, $\lambda$, with $\xb \sim \lambda$ being the typical size of a galaxy in full rotational support. If a galaxy with velocity dispersion $\sigma \sim \vvir$ (see Section~\ref{sec:model:idea:assumptions}) is in a shallow potential ($\vc(\xb)/\vvir < 1$; red color), its size will expand until the scale on which this $\sigma$ can be contained, i.e., until $\xb$ at which $\vc(\xb)/\vvir \sim 1$ (see Section~\ref{sec:model:transition}). Conversely, if the potential is steep ($\vc(\xb)/\vvir > 1$), the galaxy size will approach $\xb \sim \lambda$ at which the gravity is balanced by the centrifugal force. These two attractors are indicated with the thick purple line corresponding to ``non-disks'' and ``disk'' parts, respectively. {\bf Bottom:} The transition between non-disks and disks depends on the halo concentration, so that for higher $\cnfw$, the fraction of baryons $\fb$ needed to produce the steep potential becomes smaller, and these baryons need to be concentrated on a smaller scale. This relation is shown by the purple line, which traces the turnover point on the critical line $\vc(\xb)/\vvir = 1$, shown for a set of $\cnfw$ values. This relation enables one to formulate a criterion for disk formation with non-disks exhibiting low $\cnfw$, low $\fb$, and large $\xb$, while disks correspond to high $\cnfw$, high $\fb$, and small $\xb$ (see Section~\ref{sec:model:criterion}). 
}
\end{figure}

The above model for a galaxy embedded in an NFW halo is fully parametrized by three parameters: galaxy mass $\fb$, size $\xb$, and halo concentration $\cnfw$, enabling us to calculate $\vc/\vvir$ at $\xb$, which distinguishes potentials that are sufficiently steep ($\vc/\vvir>1$) and too shallow ($\vc/\vvir<1$) to host disks according to our model. This quantity is shown in the top panel of Figure~\ref{fig:fbxb-model} as a function of $\xb$ and $\fb$ at fixed $\cnfw = 5$ (typical for $z>2$ halos). The red color indicates the range of $\fb$ and $\xb$ at which $\vc/\vvir < 1$, implying that the potential is shallow and a galaxy with $\sigma \sim \vvir$ will be dispersion dominated, while blue indicates the region where $\vc/\vvir > 1$ and a rotationally supported disk can exist. 

Galaxies populate only a portion of this parameter space, where they can remain in a quasi-equilibrium determined by the interplay between gravity and cooling on the one hand and the support by turbulence and centrifugal forces on the other (see Section~\ref{sec:model:idea:general}). 
When the initial $\xb$ is large (e.g., close to the half-mass radius of the NFW halo shown with the horizontal dotted line), it will contract as the gas will cool and fall into the gravitational potential. If $\xb$ is in the region with $\vc/\vvir < 1$ (red), turbulent support will dominate, and the size will expand, until $\xb$ reaches the scale on which turbulent motions are contained ($\vc \sim \sigma \sim \vvir$, indicated by the ``non-disks'' branch of the purple line). If $\xb$ is below the red horizontal line (i.e., if $\xb < \lambda$), the centrifugal force will cause $\xb$ to increase until it reaches full rotational support with $\xb \sim \lambda$ (indicated by the ``disks'' branch of the purple line).
\newtext{The typical regions occupied by galaxies of different masses and at different redshifts are discussed in more detail in Sections~\ref{sec:results:mw-analogs} and~\ref{sec:results:disk-population}. In brief, MW progenitors typically have $\fb \sim 0.1$--$0.3$ and $\xb$ evolving from $\sim 0.2$ at high redshift to $\sim 0.03\ ( \approx \lambda)$ at low redshift (Section~\ref{sec:results:mw-analogs}), while lower-mass galaxies span $\fb \sim 0.01$--$0.2$, with $\fb$ decreasing toward lower halo masses (Section~\ref{sec:results:disk-population}).}

In this scenario, the transition from non-disks to disks happens when $\fb$ is large enough. This critical $\fb$ value, in turn, inversely depends on the halo concentration: a more concentrated halo contributes more mass within the region occupied by the galaxy and therefore requires a smaller fraction of baryons in that region to produce a steep potential. This is shown quantitatively in the bottom panel of Figure~\ref{fig:fbxb-model}, where the purple line traces the critical $\fb$ and the corresponding $\xb$, defined as the location of the turning point in the critical line ($\vc/\vvir = 1$) as a function of $\cnfw$. 

This purple line separates non-disk galaxies and disks. Non-disk galaxies will lie above the purple line, as, for their concentration $\cnfw$, their baryonic mass is too low (or, equivalently, their size is too large) to produce a sufficiently steep potential. Conversely, disk galaxies will occupy the region below the purple line, and will be scattered around $\xb \sim \lambda$, depending on the specific angular momentum in the galaxy relative to that of the halo and the value of $\vc/\vvir$ \citep{mo-mao-white98}.

Note that, for halos with high concentration at the moment of disk formation ($\cnfw > 10$), the region of turbulent support (colored lines) lies entirely below $\xb \sim \lambda$, which would imply that such systems always have disks. However, as we show below, such systems are not disky because they are generally hosted by low-mass halos, which are unable to counteract feedback and contain a significant fraction of baryons in the center, i.e., $\sigma > \vvir$ and the assumption of $\sigma \sim \vvir$ is violated (see Section~\ref{sec:model:idea:assumptions}). This means that, in our model, systems with $\fb \lesssim 0.1$ (i.e., where the purple line in the bottom panel intersects $\xb = \lambda$) should be treated with caution (see Section~\ref{sec:results:disk-population} for further discussion).

\subsection{Disk Formation Criterion}
\label{sec:model:criterion}

The relation between the critical $\fb$ and $\cnfw$ in the bottom panel of Figure~\ref{fig:fbxb-model} can be parametrized as:

\begin{equation}
\label{eq:fb-disk}
    f_{\rm b,disk} \approx 1.7\, \cnfw^{-4/3}.
\end{equation}
For a fixed concentration, this equation gives the minimal baryon mass fraction $\fb$ that needs to be contained in the galaxy, such that the resulting potential becomes steep enough to sustain a disk. 

This criterion can also be flipped and expressed as the minimal concentration that the dark matter halo needs to have to trigger disk formation in a galaxy containing a given baryon mass fraction:
\begin{equation}
\label{eq:c-disk}
    c_{\rm disk} \approx 1.5\, \fb^{-3/4}.
\end{equation}
\newtext{Equations~(\ref{eq:fb-disk}) and~(\ref{eq:c-disk}) reproduce the full numerical solution to within $\sim 10\%$ (and typically $\lesssim 5\%$) over the range $\cnfw \approx 1.6$--$50$ and $\fb \approx 0.01$--$0.9$ relevant for galaxies and their host halos across the range of masses and redshifts.}

Note that the galaxy size $\xb$ does not enter this disk formation criterion; instead, it is an output of the model: $\xb \sim \xcirc \sim \lambda$ for disks ($\fb > f_{\rm b,disk}$) and $\xb \sim \xturb$ for non-disks, set by the solution of $\vc(\xb) = \vvir$ at fixed $\fb$ (see the ``non-disk'' branch in the top panel of Figure~\ref{fig:fbxb-model}). 

Thus, the key component of the model required to predict disk formation is the galaxy mass fraction, $\fb$. This parameter encapsulates the uncertainties in gas cooling, accretion, star formation, and feedback that lead to the accumulation of baryonic mass at the halo center. This quantity is closely related to the stellar mass--halo mass relation (SHMR), with $\fb$ accounting for the total baryonic mass of the galaxy, including gas in addition to stars. Similar to SHMR, $\fb$ depends on $\Mvir$ and redshift, reflecting the above details of baryonic processes. In the next section, we parametrize this relation in a cosmological volume simulation and use it to test our model.

\section{Comparison with simulation results}
\label{sec:results}

In this section, we use the TNG50 cosmological simulation (briefly described in Section~\ref{sec:results:methods}) to test our disk formation model. To this end, we use the galaxy masses produced in TNG50 and compare our model predictions based on these masses with the simulated disk population.
Importantly, our goal is not to calibrate the model against the simulation. Instead, our goal is to validate the model by testing whether it can reliably predict disk formation in TNG50.

In Section~\ref{sec:results:mw-analogs}, we investigate disk formation in a sample of Milky Way (MW) analogs from TNG50 \citep{semenov23a,semenov23b} and show that our model describes the emergence of disks in such systems remarkably well. In Section~\ref{sec:results:disk-population}, we expand our analysis to the entire population of TNG50 galaxies at $z \leq 3$ and show that our model can explain the threshold halo mass for disk formation and its evolution with redshift in this simulation.

\subsection{Methods Overview}
\label{sec:results:methods}

\subsubsection{TNG50}
\label{sec:results:methods:tng50}

TNG50 is the highest resolution run of the IllustrisTNG cosmological simulation suite \citep{springel18,pillepich18b,nelson18,marinacci18,naiman18,nelson19b}. The population of galactic disks in TNG50 has been investigated in a number of recent papers \citep[e.g.,][]{pillepich19,pillepich23,sotillo-ramos22,semenov23a,semenov23b,celiz25,liang25a,liang25b}.
This section summarizes the key features of this simulation relevant to our work; for more detail, we refer to \citet{weinberger17,pillepich18,pillepich19,nelson19}.

The simulation was carried out using the quasi-Lagrangian moving-mesh $N$-body and magnetohydrodynamic code Arepo \citep{arepo}. The initial conditions for TNG50 are a representative $\sim 50^3$ comoving Mpc$^3$ volume of the Universe, assuming Planck XIII (\citeyear{planck-xiii}) cosmology: matter density $\Omega_{\rm m} = 0.3089$, baryon density $\Omega_{\rm b} = 0.0486$, cosmological constant $\Omega_\Lambda = 0.6911$, Hubble constant $H_0 = 67.74\;{\rm km\;s^{-1}\;Mpc^{-1}}$, and perturbation power spectrum with the normalization of $\sigma_8 = 0.8159$ and slope of $n_{\rm s} = 0.9667$.
The mass resolution for dark matter in TNG50 is $4.5 \times 10^5\Msun$, while the target baryonic mass resolution is $8.5 \times 10^4\Msun$, with the actual mass of gas cells and stellar particles maintained within a factor of two from this value. The gravitational softening length for dark matter, stellar, and wind particles is 575 comoving pc at $z>1$ and 288 physical pc at $z<1$, while for gas, the softening length varies with the cell size as $\epsilon_{\rm gas} = 2.5\;r_{\rm cell}$, where $r_{\rm cell}$ is the radius of a sphere with the same volume as the Voronoi cell.

In the TNG model, the gas with densities below the prescribed star formation threshold ($n < n_{\rm sf} \sim 0.1\cc$) is subject to radiative cooling (primordial, Compton, and metal-line) and heating from a spatially uniform \citet{faucher-giguere09} UV background, which is turned on at $z=6$ and is locally attenuated according to the \citet{rahmati13} gas self-shielding model, and by local AGN sources using the approximate method described in \citet{vogelsberger13}. In the star-forming ISM ($n > n_{\rm sf}$), the gas temperature and pressure are set by the \citet{sh03} effective equation of state, with modifications described in \citet{vogelsberger13} and \citet{nelson19}.

The local star formation rate, which is used to stochastically convert gas cells into star particles, is calibrated against the observed Kennicutt--Schmidt relation from \citet{kennicutt98}. 
The feedback from young stars is modeled via the contribution of SNe type II in the effective ISM pressure \citep{sh03} and via decoupled galactic winds, with a prescribed gas mass stochastically ejected from the ISM in the form of hydrodynamically decoupled particles that then recouple with the background gas after they leave the galaxy \citep{pillepich18,nelson19}. The TNG model also includes AGN feedback, modeled in two modes based on the local Eddington-limited Bondi gas accretion rate: continuous injection of thermal energy and stochastic kinetic kicks at high and low accretion rates, respectively \citep{weinberger17}.

\subsubsection{Analyses}
\label{sec:results:methods:analysis}

In our analysis, we investigate the emergence of disks in two samples of TNG50 galaxies. 
In Section~\ref{sec:results:mw-analogs}, we focus on the population of MW analogs investigated in \citet{semenov23a,semenov23b}, which develop MW-like disks by redshift $z=0$. These galaxies were selected based on their halo mass ($0.8 < M_{\rm 200c}/10^{12}\Msun < 1.4$), SFR ($\SFR > 0.2\Msunyr$), and rotational support of the young ($<100$ Myr) stellar disk at $z=0$ ($\vrot/\sigma > 6$). These criteria resulted in 61 MW analogs. 
In Section~\ref{sec:results:disk-population}, we extend our analysis to the entire population of star-forming galaxies with $\Mvir > 10^{10} \Msun$ and $\SFR > 0.01 \Msunyr$ across multiple redshifts. As we identify the presence of a disk in this population using the distribution of young star particles with ages $<100 \Myr$ (see below), the SFR threshold ensures that the selected galaxies contain at least $\gtrsim 10$ such particles, making this definition meaningful. 

To obtain the normalized galaxy size and mass, $\xb$ and $\fb$, we calculate the radius containing half of the baryon mass bound to a given dark matter halo, $r_{\rm 1/2,b}$, and the total baryon mass within $2 \times r_{\rm 1/2,b}$, and then normalize both by $\rvir$ and $(\Ob/\Om) \Mvir$, respectively.

We have tested several definitions of diskiness and found that our results remain qualitatively similar. Our tests included the definitions based on $\vrot/\sigma$, fraction of rotational energy, $\kappa = E_{\rm rot}/E_{\rm kin}$, and the fraction of young stars at high-circularity orbits (as defined in \citealt{semenov23a}), and we have applied them to either star-forming gas (i.e., the gas with $n > n_{\rm sf} \sim 0.1\cc$) or young stars with $< 100$ Myr\newtext{; varying this age cut over $\sim 50$--$200$ Myr does not change the results, except that a lower cut removes galaxies with too few young star particles from the sample}. For the MW sample (Section~\ref{sec:results:mw-analogs}), we present our results based on $\vrot/\sigma$ of star-forming gas, with $\vrot$ defined as the median rotational velocity of star-forming cells within the galaxy, while $\sigma$ is defined as a half of the 16$^{\rm th}$--84$^{\rm th}$ interpercentile range of $v_{z}$, which corresponds to one standard deviation for a Gaussian distribution. 

For a significant fraction of TNG50 galaxies from the full sample (Section~\ref{sec:results:disk-population}), the star-forming gas can formally have moderate instantaneous $\vrot/\sigma \gtrsim 1$ values \citep[see also][]{pillepich19}. However, ISM turbulence leads to rapid changes in the direction of the net angular momentum, and a coherent stellar disk does not form. Using the $\vrot/\sigma$ of a young ($< 100 \Myr$) stellar disk instead yields a significantly clearer separation between disks and non-disks; therefore, we use that definition for our full sample. Using the same definition for our MW sample does not change the results.

To investigate the connection between disk formation and the properties of the host dark matter halo, while also avoiding the back reaction of these properties to disk formation (see the Introduction), we use the halo properties from the dark-matter-only counterpart of TNG50. The halos in the two runs were matched using the {\tt LHaloTree} algorithm \citep{nelson15-illustris}.
The NFW concentrations were taken from \citet{anbajagane22}, and converted from the original 200-critical to the virial-overdensity values as $\cnfw \equiv (\rvir/r_{\rm 200c}) c_{\rm 200c}$.

\subsection{Formation of Milky Way-mass Disk Galaxies}
\label{sec:results:mw-analogs}

\begin{figure}
\centering
\includegraphics[width=\columnwidth]{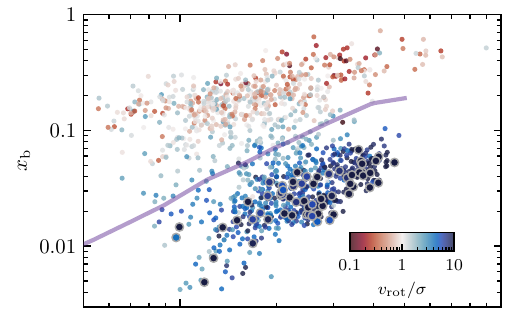}\\
\includegraphics[width=\columnwidth]{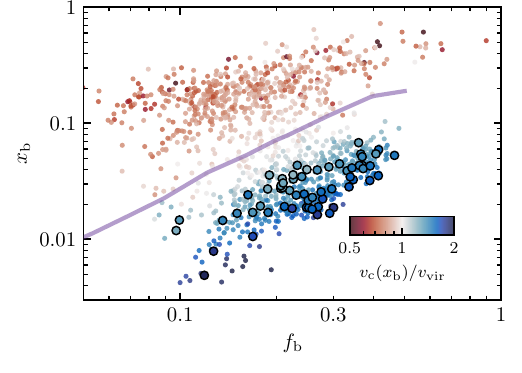}
\caption{\label{fig:fbxb-tng50mw} Evolution of the MW-like galaxies from TNG50 in the plane of the baryon half-mass size, $\xb$, and galaxy mass fraction, $\fb$. Large circles show the final locations of these galaxies at $z=0$ (when these analogs are identified; see Section~\ref{sec:results:methods:analysis}), while small circles show their locations in preceding snapshots out to $z \sim 12$. The points are colored according to the rotational support of the star-forming disk ($\vrot/\sigma$; top panel) and the potential steepness parameter ($\vc(\xb)/\vvir$; bottom panel). The purple line shows the transition from non-disks (above the line) to disks (below the line) predicted by our model (see Figure~\ref{fig:fbxb-model}). Despite the drastic differences in mass assembly histories, disk-formation timing, and the presence of destructive galaxy mergers \citep[see][]{semenov23a}, our model describes disk formation in this galaxy sample remarkably well. The comparison of colors between the two panels shows that the disk formation is associated with the steepening of the gravitational potential in agreement with our formal criterion (see Section~\ref{sec:model:criterion}).}
\end{figure}

We start from investigating the emergence of disks in the population of 61 MW-like galaxies from TNG50 \citep[see Section~\ref{sec:results:methods:analysis} and][for sample details]{semenov23a,semenov23b}.
Our model described in Section~\ref{sec:model} predicts that the presence of disks can be determined by the location of a galaxy in the plane of its total baryon mass (including stars and gas) and size, normalized to the virial mass and radius, respectively, $\fb$ and $\xb$ (recall Figure~\ref{fig:fbxb-model}). The evolution of our MW-like galaxy sample in the $\fb$--$\xb$ plane is shown in Figure~\ref{fig:fbxb-tng50mw}, with large circles showing their final location at $z=0$ and small dots corresponding to earlier available snapshots. 

The color in the top panel shows the rotational support of the star-forming gas ($\vrot/\sigma$; see Section~\ref{sec:results:methods:analysis}). The clear gradient of $\vrot/\sigma$ demonstrates that MW-like galaxies become more disky as their $\xb$ becomes smaller, i.e., as they become more compact compared to the $\rvir$ of their host halo.
As the bottom panel shows, this gradient in $\vrot/\sigma$ reflects that of our potential steepness parameter, i.e., $\vc/\vvir$ evaluated at the galaxy half-baryon-mass radius $\xb$: more disk-dominated galaxies with $\vrot/\sigma > 1$ reside in steep potentials with $\vc(\xb)/\vvir > 1$, while non-disk galaxies with $\vrot/\sigma < 1$ occupy shallow potentials with $\vc(\xb)/\vvir < 1$.

The violet line in Figure~\ref{fig:fbxb-tng50mw} shows the critical threshold in our model (see Figure~\ref{fig:fbxb-model}), and it describes the transition from non-disks to disks in our sample. As shown in \citet{semenov23a}, these galaxies exhibit a wide range of halo mass assembly histories, disk-formation timing, and the presence of galaxy mergers that are destructive to the disk, all of which lead to different tracks in the $\fb$--$\xb$ plane for individual galaxies. Despite these drastic differences, our model captures the formation of galactic disks remarkably well.

\begin{figure}
\centering
\includegraphics[width=\columnwidth]{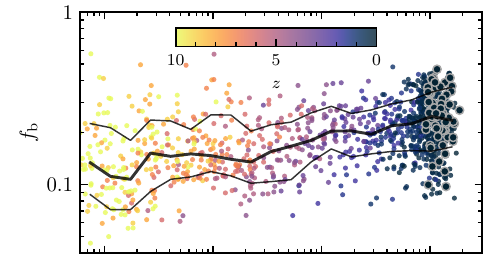}\\
\includegraphics[width=\columnwidth]{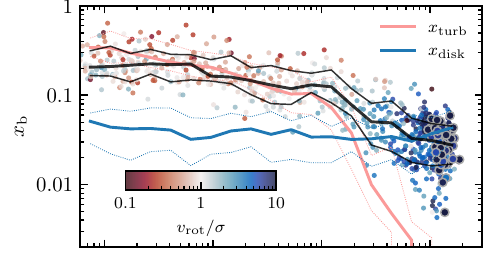}\\
\includegraphics[width=\columnwidth]{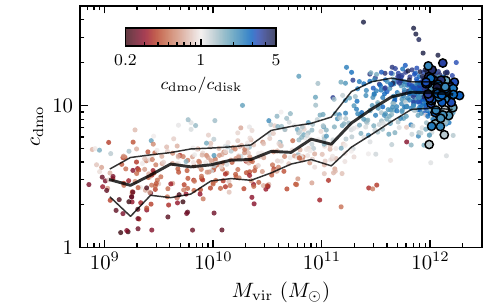}
\caption{\label{fig:fbxb-Mvir-tng50mw} The evolution of the key parameters in our model---galaxy baryon half-mass size, $\xb$, its mass fraction, $\fb$, and the dark matter halo concentration, $\cnfw$---as a function of the halo virial mass for our MW-analog sample. {\bf Top:} the evolution of $\fb$ colored by redshift. For these MW analogs, the value of $\fb$ increases only weakly with $\Mvir$ and $z$, with the typical values $\fb \sim 0.1\text{--}0.3$. {\bf Middle:} the evolution of galaxy size normalized by the virial radius, $\xb$, colored by the rotational support of the star-forming disk. Blue lines show the median and 16--84 interpercentile range of the halo spin parameter $\lambda$ as a proxy for the galaxy size in rotational support, $\xdisk$; red lines show the scale at which velocity dispersion $\sigma \sim \vvir$ would be contained by the gravitational potential, $\xturb$ (see the text for details). Disks form at $\Mvir \sim 2\times 10^{11} \Msun$, which are reached at $z \sim 2$ (see the top panel). In agreement with our model (Figure~\ref{fig:fbxb-model}), at low $\Mvir$, the sizes of dispersion-dominated systems are close to $\xturb > \xdisk = \lambda$, while at high $\Mvir$, $\xb$ settles at $\sim \xdisk$. {\bf Bottom:} the evolution of halo concentration colored by its ratio with the critical $c$ for disk formation (Equation~(\ref{eq:c-disk})). To exclude the effect of baryons on $\cnfw$, we use the values from the dark-matter-only counterpart of the TNG50 run ($c_{\rm dmo}$). The transition to disks at $\Mvir \sim 2\times 10^{11} \Msun$ occurs as a result of increasing $\cnfw$ in agreement with Equation~(\ref{eq:c-disk}).}
\end{figure}

To visualize the average evolution of these MW-like galaxies in the $\fb$--$\xb$ plane, the top two panels of Figure~\ref{fig:fbxb-Mvir-tng50mw} show the dependence of $\fb$ and $\xb$ on the halo mass, colored by redshift and rotational support, respectively. These galaxies exhibit a relatively narrow range of $\fb \sim 0.1\text{--}0.3$. Interestingly, the median $\fb$ only weakly increases at low redshift, even though the halo masses increase by three orders of magnitude, meaning that the galaxy and halo masses increase roughly self-similarly. As the figure shows, these galaxies transition to disks when their halo mass becomes $\Mvir \gtrsim 2 \times 10^{11} \Msun$ (consistent with previous studies summarized in the Introduction), which, based on the color in the top panel, occurs on average around $z \lesssim 2$.

The middle panel shows that, consistent with our model, the decrease in galaxy size results from disk formation. At early times (low $\Mvir$), the galaxy sizes are large, $\sim 0.1\text{--}0.3\,\rvir$, significantly larger than $\lambda \rvir \sim 0.04\, \rvir$ shown with the blue lines. Instead, the galaxy size is set by the turbulence containment radius $\xturb$, at which turbulent motions are contained by the given potential well (shown with the red lines). At higher masses, at which disks form ($\Mvir \gtrsim 2 \times 10^{11} \Msun$), $\xturb$ drops rapidly as the potential becomes steep, and the galaxy sizes settle around $\lambda \rvir$, as expected for rotation-supported disks \citep[][]{mo-mao-white98}.

Note that $\xturb$ shown in this figure differs from the simple definition introduced in Section~\ref{sec:model}. Although our simple definition---the radius at which $\vc = \vvir$---works well for deriving the disk formation criterion, in practice, the scale on which gas with a given velocity dispersion settles in a given potential depends on the detailed interplay between accretion, turbulence driving, dissipation, and gravity \citep[see, e.g.,][]{elmegreen-burkert10}. We find that, in TNG50, this scale is better described by a related quantity: twice the radius at which $v_{\rm esc} = \sqrt{2}\vc = \vvir$, which is shown in the figure. This definition is motivated by the requirement that the escape velocity equals the expected velocity dispersion ($\sigma \sim \vvir$; see Section~\ref{sec:model:idea:assumptions}), and multiplying this radius by $2$ yields the same numerical value as the above simple definition if the low-$r$ scaling of $\vc \propto \sqrt{M/r} \propto \sqrt{r}$ expected for the NFW and Hernquist profiles ($M \propto r^2$ at small $r$) is extrapolated to $r_{\rm turb} = \xturb \rvir$. Using the simple definition introduced in Section~\ref{sec:model} does not qualitatively change our results because this scale remains larger than $\lambda$ for $\Mvir < 10^{11} \Msun$ and drops sharply at higher masses, where disks form. The detailed investigation of $\xturb$ warrants further study in the context of the sizes of dwarf irregulars and high-redshift galaxies \citep[see Section~\ref{sec:discussion:sizes} and][]{sun25}.

The difference in the galaxy size before and after disk formation stems from the two regimes shown in the top panel of Figure~\ref{fig:fbxb-model}: confinement of a turbulence-dominated system in a shallow potential before disk formation and rotational support in a steep potential. A sufficiently steep potential for disk formation can be produced by increasing either $\fb$ or the halo concentration, but given that $\fb$ in this galaxy sample changes only weakly (the top panel of Figure~\ref{fig:fbxb-Mvir-tng50mw}), their transition from non-disks to disks is primarily driven by the increase in concentration. 

This increase in the halo concentration $\cnfw$ and the fact that it triggers disk formation in this galaxy sample are demonstrated in the bottom panel of Figure~\ref{fig:fbxb-Mvir-tng50mw}. To avoid ambiguity in the causal connection resulting from the increase in $\cnfw$ caused by disk formation, we use the halo concentration measured in the dark-matter-only counterpart of the TNG50 simulation, $c_{\rm dmo}$. The color shows the ratio of $c_{\rm dmo}$ to the critical value $c_{\rm disk}$ predicted by our model (Equation~\ref{eq:c-disk}). As the figure shows, halo concentration increases as the halo becomes more massive, and at $\Mvir \gtrsim 1 \times 10^{11} \Msun$, it increases more than the minimum value required for disk formation, which is the mass at which disks form (see the middle panel).

\begin{figure}
\centering
\includegraphics[width=\columnwidth]{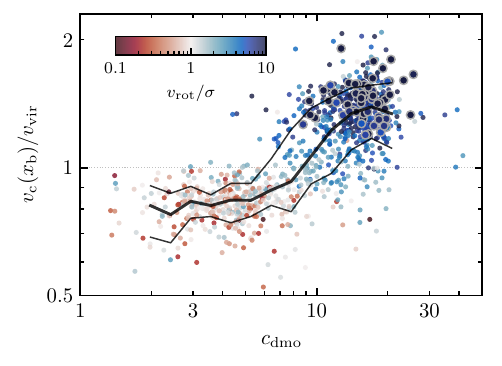}
\caption{\label{fig:c-vcvir-tng50mw} Relation between the halo concentration, potential steepness, and the rotational support (shown by color). According to our model (Figure~\ref{fig:fbxb-model} and Section~\ref{sec:model:transition}), as the galaxy mass fraction increases only weakly (Figure~\ref{fig:fbxb-Mvir-tng50mw}, top panel), the disk formation is triggered primarily by the increase of the halo concentration at large $\Mvir$ (Figure~\ref{fig:fbxb-Mvir-tng50mw}, bottom panel). When $c$ increases above 6--10, the gravitational potential for the given $\fb$ and $\xb$ becomes sufficiently steep ($\vc(\xb)/\vvir > 1$), and a disk forms ($\vrot/\sigma > 1$).}
\end{figure}

Finally, the connection between halo concentration, potential steepness, and rotational support is shown in Figure~\ref{fig:c-vcvir-tng50mw}, which demonstrates the logic behind our disk formation scenario. At early times, the halos of MW progenitors have low concentrations, $c_{\rm dmo} \lesssim 5$, which, combined with their baryonic mass fractions ($\fb \sim 0.1\text{--}0.2$), produces a shallowly rising gravitational potential with $\vc/\vvir < 1$. In such a potential, turbulent velocities are contained on a significantly larger scale than the scale of full rotational support (see Figure~\ref{fig:fbxb-Mvir-tng50mw}, middle panel), so that the system is dispersion-dominated with $\vrot/\sigma < 1$ (red color in Figure~\ref{fig:c-vcvir-tng50mw}). At later times, as dark matter halos grow, their concentrations increase and eventually become large enough to produce a sufficiently steep potential with $\vc/\vvir > 1$. In such a potential, turbulent motions are confined to a much smaller scale, and the galaxy can settle into a rotationally supported state, i.e., a galaxy disk with $\vrot/\sigma > 1$.

\newtext{We expect these trends to be general rather than specific to the TNG feedback model. The growth of $\cnfw$ that drives the transition is set by collisionless structure formation, and we use the dark-matter-only concentrations, so it is insensitive to the baryonic physics; consistently, a similar increase of the disk-formation threshold with redshift is found in the EAGLE simulation, which uses very different feedback (Section~\ref{sec:results:disk-population} and Figure~\ref{fig:Mvir-disk}). The baryonic mass fraction $\fb$, by contrast, depends on the feedback model and can produce variations between simulations (Section~\ref{sec:results:disk-population}); a detailed cross-simulation comparison is left to future work.}

\subsection{Formation of Disks across Cosmic Time}
\label{sec:results:disk-population}

\begin{figure*}
\centering
\includegraphics[width=0.32085\textwidth]{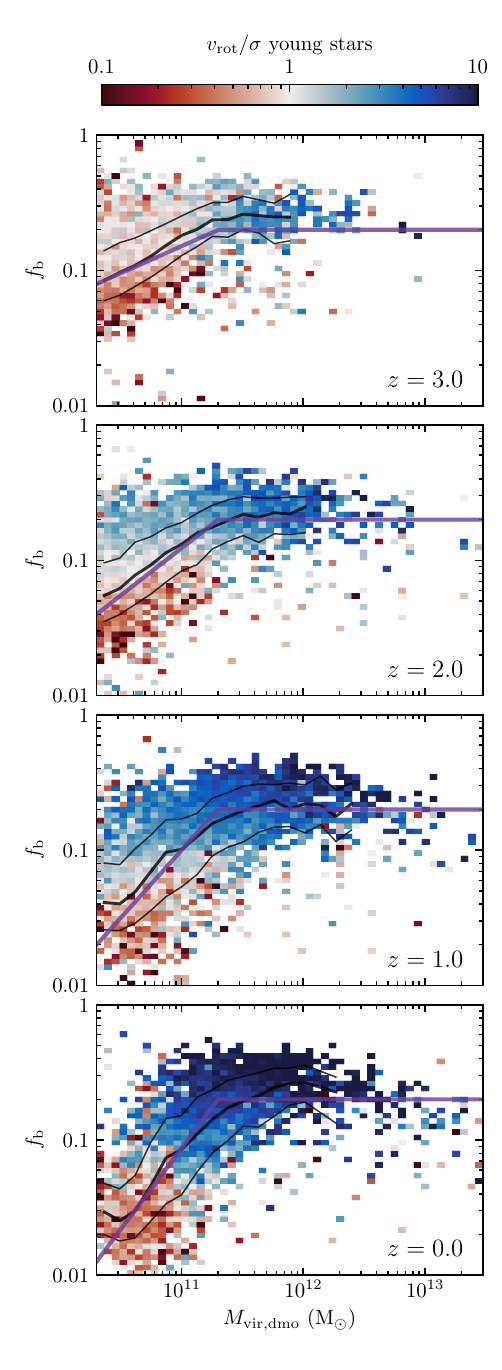}%
\includegraphics[width=0.567\textwidth]{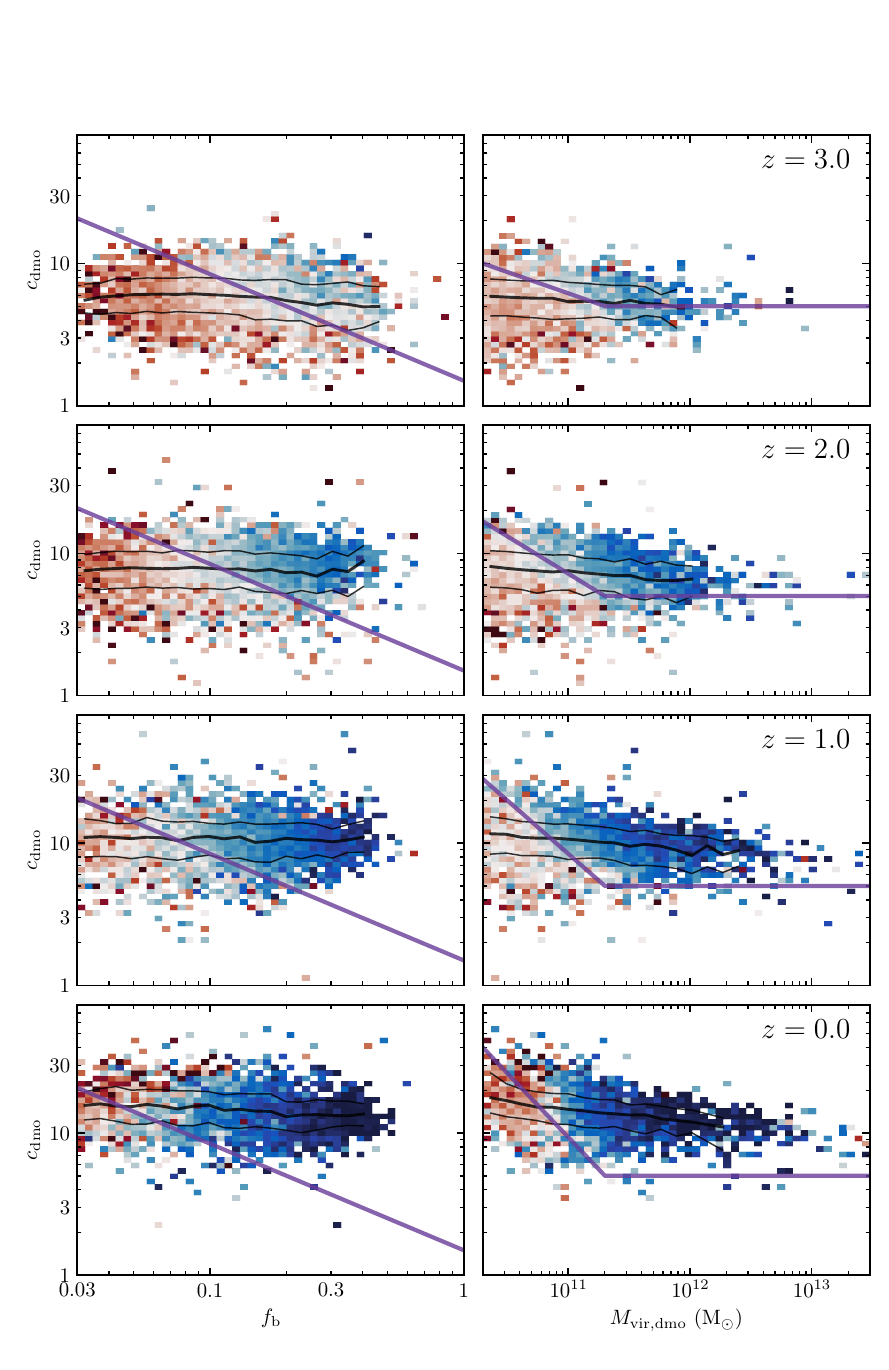}
\caption{\label{fig:tng50} The dependence of disk presence (color) on the baryonic mass fraction of galaxies, $\fb$, and the virial mass, $\Mvir$, and concentration, $\cnfw$, of their dark matter halos for the population of star-forming galaxies from TNG50. $\Mvir$ and $\cnfw$ are taken from the dark-matter-only (dmo) version of TNG50 to avoid the effects of baryons on these properties. The rows show different snapshots: $z = 3$, 2, 1, and 0, and the color shows the median rotational support of the young ($<100 \Myr$) stellar disk in each pixel. {\bf Left column:} the $\fb$--$\Mvir$ relation encapsulates the effects of baryonic physics on the galaxy growth. It is the input in our model, and we parametrize it using Equation~(\ref{eq:fb-fit}) shown with the purple line. {\bf Middle column:} the relation between $\cnfw$ and $\fb$. Our model predicts that disks form when the galaxy mass fraction or halo concentration reaches a critical value shown with the purple line (Equations~(\ref{eq:fb-disk}) and (\ref{eq:c-disk})), and this prediction agrees with the TNG50 results (transition from red to blue color). {\bf Right column: } using the $\fb$--$\Mvir$ relation from the left column, our model can be translated into the halo concentration--mass relation, $\cnfw$--$\Mvir$, shown in the right column. Again, the model captures disk formation remarkably well, suggesting a redshift-dependent halo mass threshold due to the evolution of the $\cnfw$--$\Mvir$ relation.}
\end{figure*}

In the previous section, we showed that our model successfully describes disk formation in the population of MW-mass disk galaxies.
In this section, we apply our model to the full distribution of star-forming galaxies from TNG50 across dark matter halo masses and redshifts. The goal of this section is to tie disk formation to the properties of dark matter halos predictable from the large-scale structure formation alone (such as halo masses and concentrations), to understand in which dark matter halos the transition to disks occurs, and how this transition evolves with redshift.

Figure~\ref{fig:tng50} shows the joint distributions of the baryonic mass fraction in the galaxy, $\fb$, the virial mass of its host dark matter halo, $\Mvir$, and the halo concentration, $\cnfw$, at four redshifts (from top to bottom, $z=3$, 2, 1, and 0). Again, to make sure that dark matter halo properties are not affected by baryons, we use $\Mvir$ and $\cnfw$ for the corresponding halos from the dark-matter-only version of TNG50 (see Section~\ref{sec:results:methods:analysis} for more details).

The galaxy baryonic mass fraction, $\fb$, shown in the first column, is not predicted by our model, but is rather treated as the input which we parametrize using simulation results. The relation between $\fb$ and $\Mvir$, or the galaxy baryonic mass--halo mass relation, encapsulates the effects of cooling, feedback, and other baryonic processes on the growth of the galaxy mass, causing the steepening of the potential in the halo center (see Section~\ref{sec:model:criterion}). As the figure shows, the shape of this relation remains qualitatively similar across the redshifts shown. At low $\Mvir$, $\fb$ increases with $\Mvir$, with the trend becoming shallower at higher redshift. This redshift dependence is stronger than that of the stellar mass--halo mass relation (SHMR), which includes only the stellar component \citep[e.g.,][]{behroozi19}. At $\Mvir \gtrsim 3\times 10^{11} \Msun$, the median fraction saturates at $\fb \sim 0.2$ with $\sim 0.2$ dex of scatter. At even higher masses, $\fb$ starts decreasing (similar to SHMR), but more and more of the galaxies become quiescent and therefore disappear from our sample. For simplicity, we neglect this decrease at high masses and approximate the dependence of $\fb$ on $\Mvir$ and redshift using a simple fit shown with the purple lines in the figure: 

\begin{equation}
\label{eq:fb-fit}
    \fb = 0.2 \times {\rm min} \left[ \left( \frac{\Mvir}{2\times10^{11}\Msun} \right)^{\alpha(z)} , 1 \right],
\end{equation} 
with the slopes $\alpha(z)$ being 0.4, 0.7, 1, and 1.2 at redshifts 3, 2, 1, and 0, respectively. 

The middle column of Figure~\ref{fig:tng50} shows the relation between the halo concentration and $\fb$. In TNG50, disks start to dominate at sufficiently high values of $\fb$ and $\cnfw$ (the blue region of the histograms). The purple line shows the critical line for disk formation predicted by our model ($c_{\rm disk}(\fb)$; Equation~(\ref{eq:c-disk})), which, despite the model simplicity, captures this transition remarkably well. In particular, the model elucidates the redshift dependence of the disk formation: the range of $\fb$ does not change significantly between $z \sim 3$ and 0, while halo concentrations increase at lower $z$, resulting in a larger fraction of galaxies satisfying the criterion $c > c_{\rm disk}$, which implies that their gravitational potentials become sufficiently steep to sustain disks.

Finally, the right column of Figure~\ref{fig:tng50} shows the trend of disk formation as a function of halo mass and concentration, i.e., the quantities resulting from the large-scale structure formation alone, independent of baryonic effects. The critical line shown in purple combines the relation between $\cnfw$ and $\fb$ predicted by our model (shown in the middle column; Equation~(\ref{eq:c-disk})) with the relation between $\fb$ and $\Mvir$ calibrated in TNG50 (the left column; Equation~(\ref{eq:fb-fit})). Again, our model captures the transition from non-disks to disks as a function of $\Mvir$ and redshift remarkably well. It shows that disks form in sufficiently massive halos, with the minimal halo mass for disk formation becoming larger at higher redshifts, as the average concentration at these $\Mvir$ decreases. \newtext{Our model reproduces the median location of this transition; its finite width at a given $\Mvir$ reflects the scatter in $\cnfw$ and $\fb$, as well as galaxy-to-galaxy variations in the accretion and distribution of angular momentum and in feedback, which we discuss in Section~\ref{sec:discussion:disks}.}

\begin{figure}
\centering
\includegraphics[width=\columnwidth]{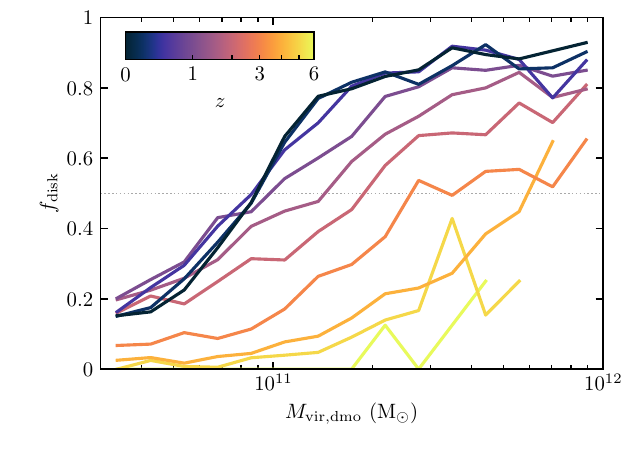}
\caption{\label{fig:fdisk} The fraction of disk galaxies (by count) as a function of virial mass at different redshifts. At fixed $\Mvir$, this fraction decreases at higher $z$, and the halo mass at which the disk fraction becomes $>50\%$ increases. The evolution of this mass with redshift is shown in Figure~\ref{fig:Mvir-disk}. As in Figure~\ref{fig:tng50}, $\Mvir$ is cited for the halos matched to the dark-matter-only version of TNG50.}
\end{figure}

This trend of the minimal halo mass for disk formation, increasing at higher redshift, is shown in Figure~\ref{fig:fdisk}. The lines in the figure show the fraction of disk galaxies with $\vrot/\sigma > 2$ (by count) as a function of halo mass. In agreement with the above results, the halo mass at which disks become dominant increases with redshift up to $z \sim 6$, when no star-forming disks satisfying our criteria remain in the simulation.
The fraction of disks depends on the choice of the $\vrot/\sigma$ threshold, with larger $\vrot/\sigma$ selecting more prominent disks. We varied this value between $\vrot/\sigma > 1$ and 3, keeping it low to quantify the transition to disks and include objects near this transition with moderate values of $\vrot/\sigma \gtrsim 1$. Our results remain qualitatively similar for these choices of the $\vrot/\sigma$ threshold.

\begin{figure}
\centering
\includegraphics[width=\columnwidth]{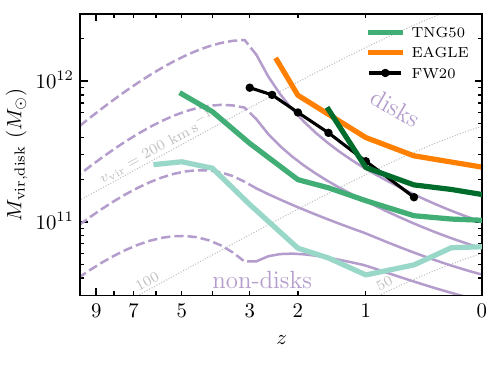}
\caption{\label{fig:Mvir-disk} Our model predictions (purple lines) can explain the increase with redshift in the critical halo mass for disk formation found in both simulations and observations. Thick green lines show the results from TNG50, specifically the halo mass at which the fraction of disk galaxies becomes $>50\%$, with the three lines showing three threshold values of $\vrot/\sigma$ to select disks, from top to bottom, $\vrot/\sigma > 3$, $2$, and $1$. The orange line shows the results from the EAGLE simulations \citep[based on][as described in the text]{trayford19}. The black line with markers summarizes the compilation of observational results by \citet{forsterschreiber-wuyts20} as detailed in the text. Our model predictions (purple lines) are based on the \citet{diemer19} concentration--mass relation and $\fb$--$\Mvir$ interpolated between $z=0\text{--}3$ (solid lines) and fixed at $z \geq 3$ (dashed lines); different lines show different constant offsets applied to the \citet{diemer19} relation, from top to bottom, 0.3, 0.15, 0, and $-0.15$ dex (see the text). Note that, because of the different definitions of disks in the shown samples, the differences in normalizations are hard to interpret. Despite these differences, all samples show an increasing trend with redshift, which our model can explain. For reference, dotted lines show the evolution of $\Mvir$ with fixed $\vvir$, which exhibits the opposite trend compared to the critical halo mass for disk formation in simulations.}
\end{figure}

The change in the critical halo mass with redshift is summarized in Figure~\ref{fig:Mvir-disk}. The green lines show the trend from TNG50, with darker colors indicating the formation of stronger disks ($\vrot/\sigma > 1$, 2, and 3). For comparison, the orange line also shows the result of the EAGLE simulation, corresponding to the lower envelope of the disk population from Figure~7 in \citet{trayford19}, with stellar masses converted to $\Mvir$ using SHMR from \citet{behroozi19}. The figure also shows the observational results summarized in Figure~4 of \citet{forsterschreiber-wuyts20}, which includes the data from \citet{kassin12,simons17,turner17,wisnioski19}. To produce the line in the figure, we use the stellar mass at which the fraction of star-forming galaxies with $\vrot/\sigma>1$ in the ionized gas becomes greater than 50\%, and convert it to halo mass using \citet{behroozi19} SHMR. 

Although Figure~\ref{fig:Mvir-disk} shows a significant difference in the normalization of the critical halo mass between different samples, this difference is hard to interpret due to the differences in disk definition. \newtext{For the observational sample,} although the threshold is defined using the same $\vrot/\sigma$ threshold as in TNG50 (for the $\vrot/\sigma > 1$ line), the large difference in the normalization may reflect the difference in the disk definition based on the kinematics of ionized gas in observations and young ($< 100$ Myr) stars in TNG50. \newtext{The ionized gas is generally more turbulent than the cold, star-forming gas from which the young stars form, with ionized gas $\sigma$ being $\sim 2\text{--}3$ times higher than molecular at $z \sim 1\text{--}2$ \citep[e.g.,][]{girard21}; beam smearing and non-turbulent motions (e.g., spiral arms) further inflate the inferred $\sigma$ in observations. Both effects lower the measured $\vrot/\sigma$ at fixed mass and shift the observational threshold toward higher halo masses, consistent with the observational line (defined by $\vrot/\sigma > 1$) falling close to our stricter $\vrot/\sigma > 2$--$3$ thresholds rather than the $\vrot/\sigma > 1$ one.}

The threshold for EAGLE, in turn, is defined by the mass at which the total contribution of prograde stellar populations becomes significant \citep[see][]{trayford19}. \newtext{Because it is based on this integrated co-rotating stellar mass rather than a per-galaxy criterion as in TNG50 and the observational sample, its normalization depends on the detailed disk and spheroid mass distributions and is not directly comparable to ours. Genuine physical differences can also contribute, given the substantially different feedback prescriptions in the two simulations. EAGLE injects a large amount of thermal energy locally \citep{dalla-vecchia-schaye12, schaye15}, which is likely more damaging to forming disks and would bias $\vrot/\sigma$ low, raising the disk formation threshold. TNG50, in contrast, employs an effective ISM model with a stiff equation of state and a decoupled wind model \citep{sh03, pillepich18}, which is more favorable to disk formation and would bias $\vrot/\sigma$ high, lowering the threshold.}
Despite these differences, the overall trend in the critical halo mass with redshift remains qualitatively similar. In both simulations and observations, the minimal halo mass for disk formation increases by almost an order of magnitude from $z = 0$ to $\sim 3$. \newtext{The similarity of the trend despite significant differences in feedback models supports the generality of our model: the trend reflects the redshift evolution of the halo concentration (see below), which is set by structure formation rather than by feedback.}

In our model, this increase in the critical $\Mvir$ is driven by the decrease of the average concentration $\cnfw$ at fixed $\Mvir$ at higher $z$. As $\cnfw$ becomes lower, a larger fraction of baryons $\fb$ needs to be contained within the galaxy to produce a sufficiently steep potential for disk formation, which requires a larger $\Mvir$ (see Figure~\ref{fig:tng50}). 

The trend predicted by our model is demonstrated quantitatively with the purple lines in Figure~\ref{fig:Mvir-disk}. To produce these lines, we adopt the concentration--mass relation, $\cnfw(\Mvir, z)$, from \citet{diemer19}, applying several constant offsets to illustrate the scatter. Using these $\cnfw(\Mvir, z)$, we define the critical $\Mvir$ as the value at which $\cnfw(\Mvir, z)$ intersects our model prediction (the purple line in the right column of Figure~\ref{fig:tng50}) and show it in Figure~\ref{fig:Mvir-disk}. The relation between $\fb$ and $\Mvir$ used in our model is interpolated between $z=0$ and 3 using Equation~\ref{eq:fb-fit}, and kept fixed at higher $z$ using the relation from $z=3$ (the model predictions are shown with dashed lines in this extrapolated regime). We find that $\cnfw(\Mvir, z)$ in TNG50, based on the concentrations reported by \citet{anbajagane22}, is $\sim 0.15$ dex higher than the \citet{diemer19} relation, which can be due to the differences in the fitting procedure and the halo finders used to calculate $\cnfw$. Given that this offset is comparable to the scatter of $\cnfw(\Mvir, z)$ at fixed $\Mvir$, we show our predictions for the \citet{diemer19} relation with four different offset values: $-0.15$, 0, 0.15, and 0.3 dex. This demonstrates both the effect of $\cnfw(\Mvir, z)$ scatter and possible additional offsets driven by, e.g., the details of halo finders, the density profile fitting procedures, the deviations from the spherical symmetry, and other uncertainties related to the halo structure. 

Our model can explain the trend of increasing minimal halo mass for disk formation with redshift out to $z \sim 3\text{--}5$. Note that this trend is the opposite of the trend of the halo mass with constant $\vvir$ (gray dotted lines). Such a trend would be expected if the disk formation threshold were set by the minimal halo mass required to contain a certain constant velocity dispersion, driven by feedback \citep[analogous to][see Section~\ref{sec:model:idea:general}]{dekel86}. Explaining the simulation trends in such a model would require that the turbulent velocity driven by feedback in a halo with a given mass decreases rapidly at low $z$. Our model, instead, suggests that the increase of the critical halo mass from $z=0$ to $z \sim 3\text{--}5$ can be explained by the evolution of halo concentrations. 

At higher redshifts, $z \gtrsim 5$, the simple extrapolation of the $\Mvir$--$\fb$ relation from $z=3$ predicts that the trend turns around and critical $\Mvir$ starts decreasing (dashed purple lines). This is because of the turnover in the redshift dependence of $\cnfw(\Mvir, z)$ at the relevant range of $\Mvir \sim 10^{11}\text{--}10^{12} \Msun$, which first decreases as $z$ increases from 0 to $\sim 3$, and then starts increasing at higher $z$ \citep[see][]{diemer19}. This trend is hard to test with TNG50, as the fraction of halos with $\Mvir$ and $\cnfw$ relevant for disk formation decreases rapidly due to the relatively small volume of the simulation, while with the bigger boxes (e.g., TNG100 and TNG300) the effects of resolution on disk formation become stronger. More importantly, the relation between $\fb$ and $\Mvir$, which is the key uncertain component of our model, is highly sensitive to the details of ISM, star formation, and feedback modeling \citep[see, e.g.,][]{semenov24a}. We leave the detailed investigation of disk formation at high $z$ to future work.

\section{Discussion}
\label{sec:discussion}

Our model described in Section~\ref{sec:model} ties the emergence of disks to the steepening of the gravitational potential via the combined effect of baryonic mass accumulation in the center and the evolution of the dark matter halo concentration \citep[see also][]{hopkins23disk}. In this section, we summarize the timeline of galaxy evolution and disk formation envisioned in our model (Section~\ref{sec:discussion:timeline}) and discuss its implications for the observed diversity of disk galaxies (Section~\ref{sec:discussion:disks}). We then discuss the role of stellar and AGN feedback in this process (Section~\ref{sec:discussion:feedback}), compare our results with other models of disk formation proposed in the literature (Section~\ref{sec:discussion:models}), and close with a remark on galaxy sizes (Section~\ref{sec:discussion:sizes}).

\subsection{The Timeline of Disk Formation}
\label{sec:discussion:timeline}

\begin{figure*}
\centering
\includegraphics[width=0.9\textwidth]{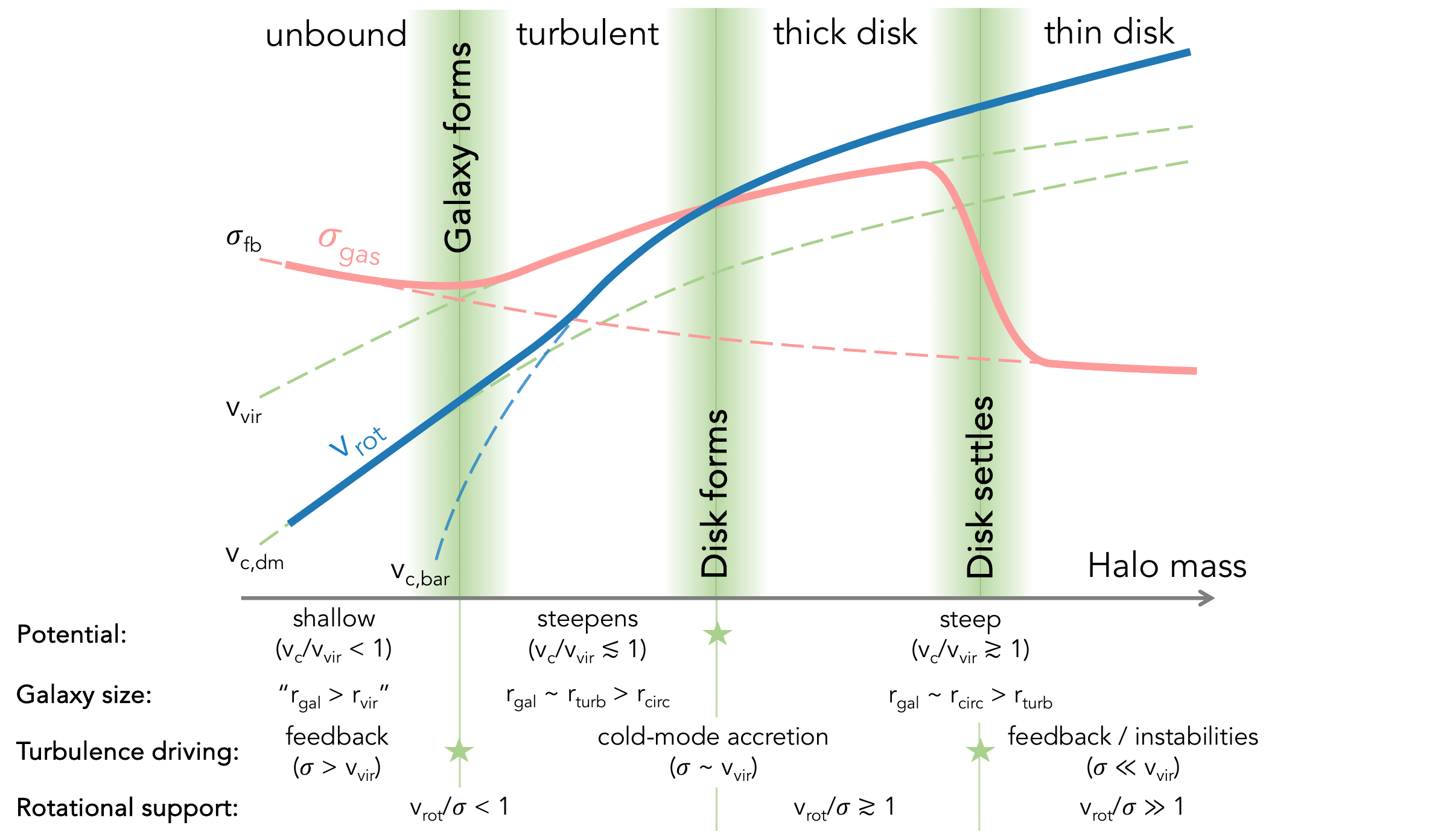}
\caption{\label{fig:disk-timeline} A schematic overview of the timeline of disk assembly. The dashed lines illustrate the qualitative evolution of the main velocity scales, such as velocity dispersions driven by feedback and gravitational instabilities ($\sigma_{\rm fb}$), the virial velocity of the host halo ($\vvir$), and the contributions to the circular velocity from dark matter ($v_{\rm c,dm}$) and baryons ($v_{\rm c,bar}$), while solid lines indicate the rotational and turbulent velocities of the galaxy, $\vrot$ and $\sigma$. Note that the coincidence of the solid and dashed lines is meant to highlight the factors that primarily determine $\vrot$ and $\sigma$, rather than to suggest exact quantitative equality. The table below the sketch lists the key quantities characterizing the evolution, with stars indicating the quantity changes that determine the transition. Our model, developed in Section~\ref{sec:model}, focuses on disk formation, i.e., the transition from turbulence-dominated systems to thick disks in the middle of the sketch.}
\end{figure*}

To put our model in a broader context of galaxy evolution, in this section, we outline the timeline of disk formation and its subsequent settling from highly turbulent state to a dynamically cold disk. The latter transition is suggested by the increasing turbulent support of the observed disk galaxies with redshift \citep[e.g.,][]{kassin12,wisnioski15,wisnioski19,simons17}, can explain the origin of thick and thin stellar disks in the Milky Way and other nearby galaxies \citep[e.g.,][]{gilmore-reid83,yoachim06,rix-bovy13,chandra23}, and has also been extensively studied in cosmological galaxy simulations in the context of settling from bursty to steady star formation \citep[e.g.,][]{muratov15,sparre17,stern21,yu21,yu22,hafen22,gurvich22,vintergatan4}.

The sketch in Figure~\ref{fig:disk-timeline} illustrates the typical evolution track envisioned for a sufficiently massive galaxy, similar to the Milky Way or typical star-forming disks at $z = 0$. The dashed lines indicate the qualitative evolution of the key velocity scales corresponding to feedback and gravity, while the solid lines are the resulting behaviors of the rotational and turbulent velocities of the galaxy, $\vrot$ and $\sigma$. The evolution of these velocity scales, and therefore the timing of transitions between stages, differs across galaxy types (see also Section~\ref{sec:discussion:disks}). The specific quantities that change in the behavior leading to the transition between the stages are indicated by star symbols under the sketch.

The red dashed line illustrates the evolution of the characteristic velocity dispersion injected locally into the gas by feedback, $\sigma_{\rm fb}$, and, at later stages, by disk instabilities (see below). The exact evolution of $\sigma_{\rm fb}$ and its dependence on the galaxy properties depends on the uncertain details of stellar and AGN feedback coupling with the ISM turbulence. In the sketch, we assume that $\sigma_{\rm fb}$ decreases with halo mass, but all the arguments below remain unchanged if $\sigma_{\rm fb}$ stays constant or increases with mass (see also Section~\ref{sec:discussion:feedback} for the discussion of the effects of feedback).

The dashed green lines show the evolution of the dark matter halo virial velocity, $\vvir$, and its contribution to the rotation curve at the scale corresponding to the galaxy size, $v_{\rm c,dm}$. Both are increasing as the halo becomes more massive and more concentrated (see also Figure~\ref{fig:vc-nfw}). The dashed blue line shows the contribution of baryons to the rotation curve at the same scale. The thick blue line shows the total $\vc$, or the rotational velocity $\vrot$ that a galaxy would have in rotational support. Note that before disk formation, the galaxy is supported by turbulent motions, and therefore its $\vrot < \vc$, but this does not change the argument. 

\emph{Early unbound stage.} At very early times, the total mass of the system is low, and the combined potential of dark matter and baryons is too shallow to contain the gas motions injected by stellar feedback. The characteristic velocity associated with feedback processes is larger than the characteristic virial velocity of the dark matter halo, $\sigma_{\rm fb} > \vvir$, preventing gas from accumulating in the halo center. This stage ends when the halo becomes massive enough so that $\vvir \sim \sigma_{\rm fb}$. 

\emph{Galaxy formation.} At $\vvir \sim \sigma_{\rm fb}$ the potential well becomes deep enough to contain the motions induced by feedback, and the baryons start accumulating in the center (the dashed blue line). The exact halo mass and redshift at which this transition occurs depend on highly uncertain details of stellar evolution and feedback coupling with the surrounding gas at these early times and low halo masses, which set the value of $\sigma_{\rm fb}$.

\emph{Bound, turbulence-dominated stage.} As the dark matter halo continues to grow, $\vvir$ becomes larger than $\sigma_{\rm fb}$, but turbulence in the galaxy can now be driven by the cold-mode accretion, maintaining $\sigma \sim \vvir$ (see Section~\ref{sec:model:idea}). This is the regime that we consider in our model. This regime persists until enough baryons are accumulated in the center to produce a sufficiently steep potential, such that the circular velocity at the characteristic galaxy size (half-mass radius in our model) becomes $\vc \sim \vvir$, implying that the galaxy reaches the state with $\vrot/\sigma \sim 1$ (see Section~\ref{sec:model:idea}). 

\emph{Disk formation (spinup).} The moment when the potential becomes sufficiently steep ($\vc/\vvir \sim 1$), marks the transition from non-disks ($\vrot/\sigma < 1$) to disks ($\vrot/\sigma > 1$). As our model suggests, the required amount of baryons in the halo center needed to trigger this transition is lower for more concentrated halos (as they contain a larger fraction of dark matter in the center), making the disk formation dependent on the halo concentration (Section~\ref{sec:model:criterion}).

\emph{Thick disk stage.} After disk forms, the potential remains steep and can continue steepening as the disk grows. While gas velocity dispersion remains close to $\sigma \sim \vvir$, the disk stays thick and highly turbulent as, for realistic mass distributions, $\vc$ is within a factor of a few larger than $\vvir$ (e.g., $\vc/\vvir < 2$ in Figure~\ref{fig:c-vcvir-tng50mw}), and therefore is only moderately larger than $\sigma$ implying that the disk can have only moderate values of $\vrot/\sigma \gtrsim 1$. \newtext{Interestingly, recent JWST measurements of edge-on galaxies at $z \sim 1$--$3$ find predominantly thin stellar disks with little thick-disk component \citep[][]{vanasselt26}, suggesting that this turbulent thick stage is short-lived or largely precedes this epoch.}

\emph{Disk settling (cooldown).} The observed velocity dispersions in the Milky Way and other nearby star-forming galaxies are of the order of $\sim 10\kms$, significantly lower than their virial velocities $\vvir \sim 100\text{--}200\kms$, and high velocity dispersions reported at high $z$. Such a ``cooldown'' of turbulent motions in disks can be a result of the formation of a hot halo, i.e., cold-to-hot accretion mode transition, leading to the formation of a subsonic cooling flow that joins the galaxy more smoothly, thereby driving lower velocity dispersions in the ISM \citep[e.g.,][]{stern21,stern23}. The origin of such disk settling, however, remains debated and can, for example, reflect further deepening of the gravitational potential \citep[e.g.,][]{hopkins23disk} or the overall decrease in the gas accretion rate.

\emph{Thin disk stage.} As the disk continues to evolve, the turbulent velocities in the ISM are mainly driven by internal processes, such as stellar feedback and the development of gravitational instabilities. The relative roles of these processes are debated \citep[e.g.,][]{krumholz16}, and given that they are expected to drive comparable velocity dispersions, in the sketch, we simply denote them as $\sigma_{\rm fb}$. The typical velocity dispersions are $\sigma \ll \vvir \lesssim \vc$, resulting in dynamically cold, thin disks with $\vrot/\sigma \gg 1$.

\subsection{Diversity of Disks}
\label{sec:discussion:disks}

The detailed tracks of the relevant velocity scales shown in Figure~\ref{fig:disk-timeline} ($\sigma_{\rm fb}$, $v_{\rm c,dm}$, $v_{\rm c,bar}$) and how $\vrot$ and $\sigma$ of the galaxy follow these tracks can differ from galaxy to galaxy, changing the timing and even the existence of specific stages. While our model considers the average behavior, the transition from non-disks to disks as a function of, e.g., stellar mass and redshift is expected to vary galaxy-to-galaxy because of intrinsic variations in the processes that lead to disk formation. 

In our model, this variation can be caused by the scatter in baryonic mass fractions and halo concentrations. Additional scatter can be introduced by the details of baryon accretion and feedback. 
For example, a coherent accretion of aligned angular momentum can result in lower turbulent velocities $\sigma$ even in low-mass galaxies, leading to earlier disk formation than implied by the schematics in Figure~\ref {fig:disk-timeline} \citep[e.g.,][]{sales12,meng-gnedin21}. Conversely, destructive mergers can temporarily or permanently destroy a gas disk even if the conditions for disk formation are satisfied. In addition, variations in the timing and energetics of feedback events during disk formation can further increase the scatter, perturbing or destroying a newly formed disk \citep[e.g.,][see also Section~\ref{sec:discussion:feedback}]{agertz16}. 

A combination of these effects can explain why the transition from irregulars to disks in the observed galaxy populations occurs gradually. While the observed stellar mass at which nearby galaxies become predominantly disky, $\Mstar \sim 10^9 \Msun$, agrees with our model (assuming the SHMR from \citealt{behroozi19}), many of the lower-mass systems still exhibit disk morphologies \citep[e.g.,][]{conselice06,kelvin14,huertas-company16,moffett16,benavides25-lowmassdisks}. A more detailed comparison with observed disk statistics is needed to distinguish among the above sources of scatter, which would be an interesting follow-up study.

Due to its strong sensitivity to the baryon mass fraction in the halo center, the timing of disk formation is also strongly sensitive to how efficiently gas is converted into stars. This is because baryon mass locked in stars is harder to remove from the potential well than gas, which can be expelled by feedback. For example, as was shown in \citet[][]{semenov24a}, resimulation of a MW analog from TNG50 with a more detailed treatment of the cold turbulent ISM leads to a significantly more efficient early star formation and also results in significantly earlier disk formation, at $z \sim 6$ compared to $z \sim 3$ in the original TNG50 run. Although the halo assembly histories are similar in these two runs (since the same initial conditions are used), the disk forms earlier in the resimulation because the galaxy accumulates enough stellar mass at its center to trigger the transition early. 

Conversely, the dependence of the disk formation threshold mass on the halo concentration could, in principle, put interesting constraints on the nature of dark matter. However, producing a noticeable effect would require the non-CDM models to significantly modify the concentration--mass relation at halo masses of a few $10^{10}\text{--}10^{12} \Msun$. Such an effect would also be hard to disentangle from feedback and other baryonic effects on the shape and normalization of the $\fb$--$\Mvir$ relation. However, this question warrants a further investigation, especially in the context of outliers such as very early massive dynamically cold disks \citep[e.g.,][]{rowland24,wang25}\newtext{, though ALMA observations of the cold gas suggest such dynamically cold disks may be common among massive high-redshift galaxies rather than rare outliers \citep[][]{rizzo24}}.

The rapid early star formation, combined with frequent galaxy mergers at early times that lead to disk formation, implies that the steepening of the potential in this scenario is caused by the formation of a bulge-like system around which a disk subsequently forms. However, observations indicate a significant population of bulgeless, thin galaxies, which seems to contradict this scenario \citep[e.g.,][]{kormendy10,bizyaev21,haslbauer22}. In our model, such bulgeless systems can be produced in high-concentration halos, such that the initial baryon concentration needed to trigger disk formation is low and can be hidden in the population of stars formed in a disk that subsequently develops and, in turn, continues to provide the sufficiently steep potential to sustain itself \citep[see also the discussion in][]{hopkins23disk}. Such objects also need to have relatively quiescent assembly histories after disk formation, as significant mergers can destroy the existing disk and rearrange it into a bulge-like configuration. This scenario of bulgeless disk formation warrants further investigation in cosmological simulations targeted at such highly concentrated, early-forming dark matter halos, which we leave to future work.

\subsection{The Role of Feedback}
\label{sec:discussion:feedback}

The three critical factors that define disk formation are the angular momentum carried by baryons, the velocity dispersion of the gas in the galaxy, and the concentration of baryons at the center, which determines both the shape of the gravitational potential and the stability of the disk after it forms. Stellar and AGN feedback can directly affect each of these factors.

First, stellar and AGN feedback can affect the angular momentum (AM) content of the galaxy in multiple ways. They can remove the low-AM gas from the galaxy, replenish it with the high-AM gas brought from the halo outskirts by outflow recycling, and keep the disk stable by keeping its gas fraction low enough to prevent fragmentation and development of instabilities funneling material to the disk center \citep[e.g.,][]{maller02,governato04,governato10,agertz15,agertz16}. In our model, these effects are accounted for indirectly by assuming that, on average, they lead to angular momentum retention by baryons during galaxy growth. Note also that this assumption only controls the size of the galaxy in the disk-dominated regime, but not the disk formation itself, which in our model depends only on the baryon fraction in the galaxy and the concentration of dark matter halo (see Section~\ref{sec:model:criterion}).

Next, feedback also shapes the potential by regulating how efficiently gas can accumulate at the center and be converted into stars. In our model, this effect is encapsulated in the fraction of baryon mass contained in the galaxy, $\fb$. In Section~\ref{sec:results}, we have validated the model by using the relation between $\fb$ and halo mass produced by the TNG50 simulations. Note, however, that disk formation can substantially change in simulations with more detailed modeling of cold, turbulent ISM (see Section~\ref{sec:discussion:disks} and \citealt{semenov24a}). For this reason, here we focused our comparison with TNG50 at relatively low redshifts $z \lesssim 3$, at which the simulation reproduces some of the key statistical properties of observed galactic disks \citep{pillepich19}. We leave the exploration of disk formation in the early Universe using more detailed simulations for future work.

Apart from its effect on the shape of the potential, $\fb$ and feedback also control the stability of the disk after it forms. Indeed, if disks are too massive, they are subject to violent global instabilities \citep[][]{fall-efstathiou80}. Such instabilities were found to be one of the key reasons why, in early galaxy simulations, baryons catastrophically lost their angular momentum, leading to spheroid-dominated systems rather than disks (see the references in the Introduction). Inclusion of efficient feedback keeps the mass fractions of such disks sufficiently low, enabling their survival \citep[e.g.,][]{agertz16}.

Another channel by which feedback can affect the shape of the potential is heating the central concentration of dark matter via strong, feedback-induced fluctuations in the gravitational potential \citep[e.g.,][]{governato10,pontzen14,dutton19}. This mechanism was proposed as a solution to the lower-than-expected dark matter concentrations inferred for observed low-mass galaxies \citep[the ``core-cusp problem''; see, e.g., reviews by][]{deblok10,bullock17}. In our model, this effect would counteract the adiabatic contraction of the dark matter halo (Section~\ref{sec:model:profile:ac}) and, therefore, can potentially increase the critical halo concentrations required for disk formation (Equation~\ref{eq:c-disk}). TNG50 does not exhibit such a strong effect of feedback on halo concentrations, which can be a consequence of smooth star formation histories produced by the TNG model \citep[e.g.,][]{roper23}. This effect, therefore, needs to be explored in more detailed, high-resolution simulations that produce such bursty behavior.

Finally, stellar feedback can affect galaxy morphologies by directly driving turbulent motions in the ISM. 
In our model, we assume that velocity dispersion in the ISM is driven primarily by cold-mode accretion at the level of $\sigma \sim \vvir$, with feedback contributions either subdominant or comparable. By contributing to turbulence at these early stages, feedback can significantly perturb or even destroy disks at the verge of formation, affecting the timing of disk formation and prolonging the duration of the thick disk stage \citep[see also][]{kaufmann07,dalcanton10,agertz16}. 

For example, qualitative differences in feedback treatment may explain why disk formation timescales differ qualitatively between, e.g., TNG and FIRE. Feedback modeling in TNG is favorable for disk formation, as the stiff effective equation of state \citep{sh03} keeps the forming disk pressurized and stable against gravitational instabilities, while the decoupled wind model preferentially removes low angular momentum gas from the center without imparting turbulence directly into the ISM (although it can also disrupt early disks, affecting the transition; \citealt{celiz25}). MW-mass galaxies in this simulation quickly form thin disks, with distinct populations of stars reminiscent of the thick disk at $z=0$, produced mainly by heating the preexisting thin disk in mergers \citep{chandra23,semenov23b}. Comparison with observed galaxies indicates that the resulting disks are too gravitationally stable and have too little variance in the stellar Toomre-Q parameter \citep{romeo20}. On the other hand, in FIRE, stellar feedback is vigorous and bursty, imparting turbulence directly into the ISM and substantially perturbing the ISM of early galaxies. With such a model, stars at early times form in a thick turbulence-dominated disk that spins up rather gradually, while the settling into a thin disk happens rapidly \citep[e.g.,][]{yu21,yu22,gurvich22}.

\newtext{Note, however, that a stronger or more disruptive effect of feedback on disks does not simply correspond to ``more'' feedback; the outcome depends on how feedback acts rather than on its total energy or momentum budget. More efficient early feedback from radiation pressure, photoionization, and stellar winds disperses dense star-forming clumps, lowering the clustering of the subsequent supernovae and making the integrated feedback less bursty and less disruptive to a forming disk \citep[][]{smith21,keller22}. Burstiness is likewise sensitive to modeling details, such as the directional coupling of supernova momentum at fixed total momentum \citep[e.g.,][]{zhang24} and the adopted star formation threshold \citep[e.g.,][]{buck19,bose19}. Reducing ISM turbulence or burstiness therefore does not necessarily imply weaker feedback.}

\newtext{Cosmic rays (CRs) provide a further example in which stronger feedback can favor disks. When their diffusion is suppressed in star-forming regions, local CR pressure gradients inhibit clump formation and help stabilize the disk \citep[][]{semenov21-cr}, while a CR-dominated gaseous halo can facilitate disk formation by reducing accretion-driven turbulence and self-aligning the accreted angular momentum \citep[][]{hafen22}.}

\subsection{Comparison with Other Models}
\label{sec:discussion:models}

Formation of galactic disks has been the subject of active research in the past decades, and several mechanisms have been proposed as the key factors. In Section~\ref{sec:model:idea}, we have compared some of these models in terms of the differences in the interplay between the gravitational potential, gas angular momentum, and turbulence driving. To frame our model within this broader context, here we summarize how these mechanisms fit in the disk formation timeline described in Figure~\ref{fig:disk-timeline} and Section~\ref{sec:discussion:timeline}.

The idea of disk formation triggered by the steepening of the gravitational potential was proposed by \citet{hopkins23disk}. Based on a suite of galaxy formation simulations with varying parameters, the authors concluded that the presence of the disk is correlated with the steepness of the potential expressed in terms of an empirical ratio of characteristic scales describing the potential (such as the radius at which $\vc$ exhibits a specific slope or reaches its maximum) and the galaxy size (such as the radius containing a certain mass fraction of gas or the circularization radius). The authors explain this connection by a steep potential providing a geometric center for disk formation and efficient damping of radial motions of gas (breathing modes). 

In this paper, we provide a further physical motivation for why potential steepening is important for disk formation (Section~\ref{sec:model:idea}) and propose a formal criterion for the steepness required to trigger this transition. \newtext{Whereas \citet{hopkins23disk} established this connection empirically, by correlating disk presence with a steepness ratio across a suite of FIRE simulations with varying parameters, our criterion follows from a physical argument and is predictive directly in terms of the halo concentration and baryonic mass fraction, $\cnfw$ and $\fb$, respectively.} Specifically, equating the scales of centrifugal and turbulent support leads to the critical condition, such that $\vc/\vvir \sim 1$, when $\vc$ is evaluated on the scale of the galaxy, e.g., the baryon half-mass radius, \newtext{with $\vc$ carrying the dependence on $\cnfw$ and $\fb$}. This condition marks the disk formation in Figure~\ref{fig:disk-timeline}, with $\vc/\vvir < 1$ corresponding to shallow potentials that are unable to contain turbulent motions, leading to turbulence-dominated systems, while $\vc/\vvir > 1$ corresponds to disks. \newtext{Our analysis and that of \citet{hopkins23disk} also rely on different simulations, TNG50 and FIRE, which use very different feedback prescriptions and numerical methods. That two independent analyses converge on potential steepening as the driver of disk formation is itself a reassuring result, supporting the generality of this picture.}

Another critically important factor considered in the literature is the interplay between the stellar feedback and the deepening of the gravitational potential as the host halo and the galaxy itself grow in mass. As a prerequisite for disk formation, the motions imparted into the gas by feedback need to be contained within the total potential well \citep[][]{dekel86,maclow99-blowout,ferrara00,read06}. This condition sets the minimal galaxy and halo mass, below which feedback expels the available gas away from the halo, corresponding to the transition between the ``unbound'' and turbulence-dominated stages in Figure~\ref{fig:disk-timeline}. 

Note, however, that such a condition alone cannot explain the mass threshold for disk formation. Assuming that an analogous critical value of $\vvir$ is responsible for disk formation, would result in the minimal mass of dark matter halos hosting disk galaxies to increase at low $z$ as halo mass at fixed $\vvir$ increases. This is the opposite from the tracks shown in Figure~\ref{fig:Mvir-disk}. Explaining such a behavior would require the critical $\vvir$, and therefore the characteristic velocity dispersion imparted into the ISM to rapidly decrease at low redshift. In addition, as was shown by \citet{hopkins23disk}, it is possible to produce non-disk galaxies even in deep potentials as long as the potential is sufficiently shallow. Note, however, that feedback can still contribute to the velocity dispersion during the early turbulence-dominated stage, thereby affecting disk formation and the duration of the thick disk stage (see Section~\ref{sec:discussion:feedback}). 

\newtext{A related framework is the two-phase model of galaxy formation \citep{mo24}, in which the cooled gas is initially self-gravitating, turbulent, and dynamically hot, and settles into a rotation-supported disk only after the halo potential stabilizes and the central gas fraction drops. Disk formation is then controlled by the condition $f_{\rm gas} \lesssim \lambda$, so that the gas reaches rotational support before becoming self-gravitating. In terms of our framework, this condition corresponds to a decrease of $\sigma$ relative to $\vvir$ rather than to a steepening of $\vc(r)$ (the two routes discussed in Section~\ref{sec:model:idea:general}). Consistently, \citet{ma25} find in FIRE-2 simulations of MW-mass galaxies that the system is dynamically hot and bulge-dominated at early times and settles into a cold disk later, as the central gas fraction declines.}

Another factor affecting disk formation is the way how material is delivered into the galaxy, specifically the frequency of mergers and the geometry of gas inflow \citep[e.g.,][]{sales12,dekel20,meng-gnedin21}. While the differences in mergers and inflow geometries can change the timing of disk formation for individual galaxies (see also Section~\ref{sec:discussion:disks}), they do not change the results of our model, as it considers the instantaneous conditions for the formation of a gas disk. At early times, before disk formation according to our model, such mergers would contribute to driving ISM velocity dispersions at the level of $\vvir$. After disk formation, destructive mergers with misaligned spin can destroy the preexisting stellar disk; however, as the net angular momentum of the merger remnant is lower than that of the original disk \citep[e.g.,][]{barnes92,barnes96}, it will be more compact and therefore feature an even steeper potential, so that the remaining and newly accreted gas will quickly reform a disk \citep[see also][]{hopkins09}. Thus, while our model can be used to predict the possibility of a gas disk formation after such mergers, to explain the population of stellar disks, one needs to take into account galaxy mergers, the fate of gas in such mergers, and the timescales for disk reformation. 

The accretion flow can also change as a result of hot gaseous halo formation \citep[e.g.,][]{dekel-birnboim06,dekel09}, which could affect disk formation \citep[][]{stern19,stern20,stern21,stern23,gurvich22,hafen22}. However, the halo mass at which such a cold-to-hot mode transition occurs, $\Mvir \sim 10^{12} \Msun$ is an order of magnitude larger than the halo masses at which disks are reported in simulations and observation ($\Mvir \sim 10^{11} \Msun$; see Figure~\ref{fig:Mvir-disk} and references in the Introduction) and therefore cannot explain disk formation by itself. In Figure~\ref{fig:disk-timeline}, this transition is attributed to the disk settling from turbulent, thick to dynamically cold, thin disks.
Note, however, that if such a transition can be produced at lower halo masses, e.g., by establishing a cosmic ray-dominated halo, mimicking the effects of the hot halo \citep{hafen22}, it can also modify the disk formation by reducing velocity dispersions at early times. This warrants a further systematic study of disk formation in simulations with cosmic ray feedback.

Finally, it is worth noting that the duration of the transitions can be used to distinguish between different scenarios. For example, simulations of an early-forming MW analog with detailed ISM modeling by \citet{semenov24a} showed that the galactic disk develops on a very short timescale, $\sim 100\Myr$. As this timescale is comparable to the local dynamical timescales of the galaxy, e.g., its orbital period or crossing time, this suggests that it is a local process that triggers this transition. The steepening of the central potential is one such process. In contrast, changes on the scale of the entire halo, such as the transition in the accretion flow or frequency of mergers, are expected to occur on longer dynamical timescales of the entire halo.

\subsection{Note on the Galaxy Sizes before Disk Formation}
\label{sec:discussion:sizes}

One of the results of our model is that the galaxy size, enclosing the half mass of all baryons, is significantly larger than $\lambda \rvir$ expected for disks in rotational support \citep[e.g.,][]{fall-efstathiou80,mo-mao-white98}. In our model, these inflated sizes originate from the scales on which the turbulent motions are contained by a given potential being larger than the rotational support radius (see Figure~\ref{fig:schematics} and Section~\ref{sec:model:idea}). 

The elevated sizes of early, pre-disk galaxies has also been suggested by \citet{sun25} based on FIRE simulations and an analytical model for the turbulent structure of such early galaxies. The gas in early simulated halos is highly turbulent while the virial temperatures are low, implying that the turbulence is supersonic and highly compressive. This, combined with a higher average density of the early Universe, results in a situation where the conditions suitable for star formation can be satisfied within a significant portion of the halo out to much larger scales than $\lambda \rvir$. This effect complements the increase in galaxy sizes due to strong turbulent support implied by our model and can further inflate them. Accurately predicting galaxy sizes at early times requires taking both effects into account and warrants further investigation.

\vspace{2em}
\section{Summary and Conclusions}
\label{sec:summary}

In this paper, we have investigated the physical origin of the mass threshold for disk galaxy formation and its evolution with redshift. We developed a simple, physical framework that links disk formation to the structure of the gravitational potential in which galaxies evolve. 

In this picture, the galaxy morphology is governed by the competition between centrifugal support and turbulent motions.
A disk forms when the scale on which turbulence is contained by the gravitational potential becomes smaller than the scale of full rotational support, which can be achieved either at low velocity dispersions or in steep potentials (Figure~\ref{fig:schematics}). As the turbulent velocities at the early stages when disks form can be sustained by active gas accretion at the level of virial velocities of the host dark matter halo, this transition is mainly driven by the steepening of the gravitational potential \citep[Section~\ref{sec:model:idea:assumptions}; see also][]{hopkins23disk}.

By considering a simple model for the mass distribution of a galaxy at the verge of disk formation, described by a spherical concentration of baryons embedded in an NFW halo, we have derived a criterion for disk formation linking the baryonic mass fraction in the galaxy and the halo concentration (Equations~(\ref{eq:fb-disk}) and (\ref{eq:c-disk})) and tested its predictions against the results from the TNG50 cosmological simulation. Our main conclusions can be summarized as follows:

\begin{enumerate}
    \item The steepness of the potential in the halo center required for disk formation is provided by the mass contributions from both the galaxy itself and the dark matter halo. As a result, the galaxy mass required to trigger the transition decreases at higher halo concentration (Equation~\ref{eq:fb-disk}). Alternatively, this relation can be interpreted as the minimal halo concentration needed to trigger disk formation at a fixed galaxy mass (Equation~\ref{eq:c-disk}).

    \item The ratio of the galaxy size (accounting for both stars and gas) to the virial radius changes systematically as a result of disk formation. The classical relation, where this ratio is set by the halo spin parameter, holds only when a rotationally supported disk is established. Before disk formation, the size is bigger and is set by the scale on which turbulent motions are confined in the potential (Figures~\ref{fig:fbxb-model} and \ref{fig:fbxb-Mvir-tng50mw}, middle panel).

    \item Our model introduces a formal definition of the steepness of the potential required for disk formation via the ratio of the circular velocity measured at the galaxy's baryon half-mass radius to the virial velocity of the host halo, $\vc/\vvir$. The potential is sufficiently steep to sustain a disk when this ratio is $\vc/\vvir > 1$. Formation of disks in a sample of MW-mass galaxies from TNG50 agrees well with this criterion (Figures~\ref{fig:fbxb-tng50mw} and \ref{fig:c-vcvir-tng50mw}).

    \item In the sample of TNG50 MW analogs, the baryonic mass fraction contained in the galaxy evolves only weakly, and the potential steepening leading to disk formation results from the increase of the halo concentration as the halo grows in mass. These results are based on the halo concentrations from the dark-matter-only version of TNG50, which are not affected by disk formation, therefore suggesting that the increase in concentration triggers disk formation, not the other way around (Figures~\ref{fig:fbxb-Mvir-tng50mw} and \ref{fig:c-vcvir-tng50mw}).

    \item We have also tested our model against the evolution of the entire population of star-forming galaxies in TNG50 between $z \sim 0\text{--}3$. Assuming the relation between the galaxy total baryon mass and the host halo mass, $\Mvir$, produced in the simulation, our model predicts the critical concentration needed for disk formation as a function of $\Mvir$. The transition to disk-dominated systems in TNG50 agrees well with this relation (Figure~\ref{fig:tng50}).

    \item The model predicts that the typical halo mass at which disks start to form increases by roughly an order of magnitude from $z \sim 0$ to 3 (Figure~\ref{fig:fdisk}). This is because the average halo concentrations decrease at higher redshift, thereby shifting the transition to larger $\Mvir$, which contain a larger mass fraction in the galaxy and therefore require a lower concentration for disk formation (Figure~\ref{fig:tng50}). The resulting redshift evolution of the critical halo mass qualitatively agrees with the results of other simulations and observations (Figure~\ref{fig:Mvir-disk}).
\end{enumerate}

One of the uncertainties in existing models of disk formation was establishing the causal connections among the various factors that might affect this transition. For example, in simulations, significant transformations in the accretion flow and potential shape occur simultaneously with disk formation, making such a connection hard to pinpoint (see references in the Introduction). In our comparisons with TNG50, we mitigate this uncertainty by using the halo properties from the dark-matter-only run. The agreement with our analytic model, therefore, suggests that it is the change in halo concentration and the associated steepening of the potential that leads to disk formation. Other physical factors, such as the confinement of feedback-driven turbulence in the potential and cold-to-hot accretion mode transition, can still play significant roles in other stages of galaxy evolution, such as thick-to-thin disk transition (Figure~\ref{fig:disk-timeline} and Section~\ref{sec:discussion:timeline}). 

Finally, the predicted evolution of the disk formation threshold depends on the total baryon mass of the galaxy, as obtained from the TNG50 simulation. For this reason, the agreement of our model with the simulation results should be thought of as validation of the model: our model can successfully explain disk formation in the simulation by adopting the total baryon mass--halo mass relation predicted in that simulation. However, the evolution of this relation and disk formation are strongly sensitive to the details of ISM, star formation, and feedback modeling, in particular, in the extreme regimes of the early Universe (see \citealt{semenov24a,semenov24b} and Section~\ref{sec:discussion:disks}). Therefore, in future work, it will be interesting to investigate more detailed models of these early stages of galaxy evolution to shed light on the formation of the first galactic disks now observed by JWST and ALMA.

\section*{Acknowledgements}
\newtext{I thank the referee for the insightful comments that helped to improve this manuscript.}
I am also deeply grateful to Andi Burkert, Charlie Conroy, Lars Hernquist, Phil Hopkins, Andrey Kravtsov, and Volker Springel for insightful discussions. 
The analyses presented in this paper were performed on the FASRC Cannon cluster supported by the FAS Division of Science Research Computing Group at Harvard University.
My support was provided by Harvard University through the Institute for Theory and Computation Fellowship. 
I am also grateful to the Excellence Cluster ORIGINS for the hospitality and travel support through the Deutsche Forschungsgemeinschaft (DFG, German Research
Foundation) under Germany's Excellence Strategy -- EXC-2094 --
390783311.
Analyses presented in this paper were greatly aided by the following free software packages: {\tt yt} \citep{yt}, {\tt NumPy} \citep{numpy_ndarray}, {\tt SciPy} \citep{scipy}, {\tt Matplotlib} \citep{matplotlib}, {\tt COLOSSUS} \citep{colossus}, and \href{https://github.com/}{GitHub}. We have also used the Astrophysics Data System (\href{http://adsabs.harvard.edu/abstract_service.html}{ADS}) and \href{https://arxiv.org}{arXiv} preprint repository extensively during this project and writing of the paper.

\bibliographystyle{aasjournalv7}
\bibliography{}

\end{document}